\documentclass[12pt,preprint]{aastex}
\usepackage{natbib}
\usepackage{graphicx}
\newcommand{\nh}{$N_{\mathrm H}$}

\begin{document}

\title{The Complex X-ray Spectrum of the Sefyert 1.5 Source NGC 6860}
\author{Lisa M. Winter\altaffilmark{1,3} \and Richard Mushotzky\altaffilmark{2}}

\altaffiltext{1}{Center for Astrophysics and Space Astronomy, University of Colorado, Boulder, CO, USA}
\altaffiltext{2}{Astronomy Department, University of Maryland, College Park, MD, USA}
\altaffiltext{3}{Hubble Fellow}

\begin{abstract}
The X-ray spectrum of the Seyfert 1.5 source NGC 6860 is among the most complex of the sources
detected in the Swift Burst Alert Telescope all-sky survey.  A short XMM-Newton follow-up
observation of the source revealed a flat spectrum both above and below 2\,keV.  To
uncover the complexity of the source, in this paper we analyze both a 40\,ks Suzaku and a 100\,ks XMM-Newton observation of NGC 6860.  While the spectral state of the source changed between the newer observations presented here and the earlier short XMM-Newton spectrum -- showing a higher flux and steeper power law component -- the spectrum of NGC 6860 is still complex with clearly detected warm absorption signatures.  We find that a two component warm ionized absorber is present in the soft spectrum, with column densities of about $10^{20}$ and $10^{21}$\,cm$^{-2}$, ionization parameters of $\xi = 180$ and 45 ergs cm s$^{-1}$, and outflow velocities for each component in the range of $\approx 0$--300 km\,s$^{-1}$.  Additionally, in the hard spectrum we find a broad ($\approx 11000$\,km\,s$^{-1}$) Fe K$\alpha$ emission line, redshifted by $\approx 2800$\,km\,s$^{-1}$.  
\end{abstract}

\section{Introduction}
Through the very hard X-ray survey (14 -- 195\,keV) conducted by Swift's
Burst Alert Telescope (BAT), a sample of unbiased towards absorption (for
\nh~$< 10^{24}$\,cm$^{-2}$), local ($<$z$> \approx 0.03$), AGN have
been identified \citep{2008ApJ...681..113T}.  Among these 153 sources, we obtained
10\,ks XMM Newton follow-ups for 22 previously unstudied, in the X-ray band, BAT AGNs.  These
sources, selected purely on their significant detection ($> 4.8 \sigma$) in the BAT and
probable optical identification with bright sources in the 2MASS/Digital Sky Survey,
comprise a representative sample of the BAT AGNs, with half of the sources showing low X-ray
column densities (\nh~$ < 10^{23}$\,cm$^{-2}$) and half more heavily absorbed \citep{2008ApJ...674..686W}.  From our analysis of both these XMM Newton follow-ups and a combination of published spectra and our analyses of Swift XRT/ASCA/XMM Newton spectra of the remaining sources in the 9-month catalog \citep{2009ApJ...690.1322W}, we found that the majority of the X-ray spectra
were well described by simple models, such as an absorbed power law, partial covering, soft excess, warm absorber (through fits with \ion{O}{7} and \ion{O}{8} edges), or reflection.  Such simple models can even be applied to the
more complicated spectra of Compton-thick sources, where a strong Fe K$\alpha$
fluorescent line and flat power law spectrum indicates a reflection-dominated spectrum.

However, such simple models could not fit the 0.3--10\,keV spectrum of roughly 10\% of the BAT-detected AGN, including NGC 6860. 
Unlike any previously known sources, the spectrum of this source was flat ($\Gamma < 1.0$)
both below and above 2\,keV, no matter how we modeled the continuum.   To our knowledge, there is no other known AGN to exhibit a flat spectrum below 2\,keV.  Likely, this flat spectrum at low energies is due to complex line structure that is unresolved in the $\approx 10$\,ks MOS spectra (the pn spectrum unfortunately could not be extracted due to problems with the observation data files (ODFs)).  Further, we found that the spectrum of this source could be explained by a number of models, in particular a warm absorber or double partial covering model fit equally well.  This degeneracy could not even be broken with the inclusion of the 14--195\,keV BAT spectra.  

To unravel the complexities of this unusual source, we obtained a long XMM-Newton exposure including both EPIC CCD and RGS spectra.  We also include an analysis of the archived Suzaku observation of this source, which provides simultaneous coverage of the soft, hard, and very hard X-rays through the XIS and HXD PIN detectors.  NGC 6860 is a southern hemisphere ($\alpha = $20:08:46.89, $\delta = $--61:06:00.70) target optically identified as a Seyfert 1.5 source \citep{1993ApJ...405..186L}.  The source is nearby ($z = 0.014884$) and hosted in a barred spiral galaxy (SBab).  Optical emission line diagnostics reveal the source as an AGN/starburst composite, with the AGN dominating only in the inner 10$\arcsec$ in the optical/IR bands \citep{2006AA...459...55B}.  Thus, the X-ray observations will also likely include contributions from both the starburst and AGN.  In Section~\ref{observations} we describe the XMM-Newton and Suzaku observations.  In Section~\ref{var}, we include an analysis of the light curves.  In Section~\ref{spectra}, we discuss the spectral analysis.  We discuss some of the unusual properties of this source in Section~\ref{unusual}.  Finally, we summarize the results of our analysis in Section~\ref{summary}.

\section{Observations and Data Reduction}\label{observations}
\subsection{XMM-Newton Observations}
NGC 6860 was observed on 29 March, 2009 for 123\,ks (observation 0552170301).  The XMM-Newton data were reduced using the Science Analysis System (SAS) version 8.0. Calibrated photon event files were created for the EPIC pn and MOS cameras using the ODFs and the commands {\tt epchain} and {\tt emchain}, respectively.  The event tables were filtered using the standard criteria outlined in the XMM-Newton ABC Guide\footnote{See \url{http:// heasarc.gsfc.nasa.gov/docs/xmm /abc/}}.  For the MOS data, good events correspond to a pulse height in the range of 0.2-12 keV and event patterns that are characterized as 0-12 (single, double, triple, and quadruple pixel events). For the pn camera, only patterns of 0-4 (single and double pixel events) are kept, with the energy range for the pulse height set between 0.2 and 15 keV. Bad pixels and events too close to the edges of the CCD chips were rejected using the stringent selection expression ``FLAG = 0''.  

The source and background spectra, along with response and ancillary response files,  were then extracted using the SAS task {\tt especget}.  The source spectra were extracted from a circular region centered on the source, with a radius of 120\arcsec.  The background spectra were extracted from a circular region of the same size on the same chip.  The spectra were then grouped using the FTOOL {\tt grppha}, binning the spectra by 20 counts.  The resultant exposure times of the spectra were 100\,ks (pn) and 117\,ks (MOS1 and MOS2).  The corresponding count rates for the spectra are 9.5\,counts\,s$^{-1}$ (pn) and 2.6\,counts\,s$^{-1}$ (for MOS1 and MOS2).

Since the XMM-Newton spectrum of our source suffers from high background count rates (particularly during the last $\approx 8$\,hours of the exposure), we also extracted a pn spectrum from an event file filtered by count rate.  We chose the stringent filtering of keeping only events whose count rates are below 100.  The filtered pn spectrum has an average count rate of 9.8\,counts\,s$^{-1}$ and an exposure time of 66\,ks.  We compared the spectra of the filtered and unfiltered pn count rates.  We find that the only differences in spectral shape occur above about 8\,keV.  At these high energies, the unfiltered spectrum shows a curvature which is not replicated in the filtered spectrum.  The filtered high energy spectrum is instead well fit by a power law.  Throughout the text, we use the unfiltered spectrum, which has more counts and thus a higher signal-to-noise ratio, for all but the broad band spectral fits (for which we use the filtered pn spectrum).  

For the RGS spectra, we extracted first-order spectra for the RGS1 and RGS2 using the SAS command {\tt rgsproc}.  Due to flaring seen in the light curve, we created a good time interval file using the SAS task {\tt tabgtigen} and filtering out all events with count rates $> 0.2$.  The spectra were re-extracted using this filtering.  Response files were created using the task {\tt rgsrmfgen} and the spectra were rebinned to 20\,counts per bin using {\tt grppha}.  The exposure times for the RGS spectra are 93\,ks, with count rates of 0.21\,counts\,s$^{-1}$ (RGS1) and 0.25\,counts\,s$^{-1}$ (RGS2).

\subsection{Suzaku Observations}
In addition to the XMM-Newton observation, we include in our analysis the archived 44\,ks Suzaku observation of NGC 6860 (observation 703015010) from 7 April 2008, observed in HXD nominal pointing mode.  We used the cleaned version 2.0 processed event files supplied by the Suzaku team.  To extract the spectra, we combined the $3 \times 3$ and $5 \times 5$ edit modes for each of the front-illuminated (XIS0 and XIS3) and back-illuminated (XIS1) CCDs.  Source and background spectra were extracted in circular regions of 210\arcsec and 180\arcsec, respectively, in {\tt XSELECT}.  The response and ancillary response files were created using {\tt xisrmfgen} and {\tt xissimarfgen}.  We then added the spectra and response files from the front-illuminated detectors (using {\tt mathpha}, {\tt addrmf}, and {\tt addarf}) and rebinned the spectra and rmf files from 4096 to 1024 channels (which is still larger than the resolution of the CCDs).  The front-illuminated and back-illuminated spectra were each binned to 20 counts per bin.  The resultant exposure times are 44\,ks with count rates of 1.34\,counts\,s$^{-1}$ (XIS1) and 1.13\,counts\,s$^{-1}$ (the average for the combined XIS0 + XIS3 spectrum).

For the HXD spectra, we only extracted spectra from the PIN instrument.  Spectra were extracted using {\tt XSELECT} for both the source and Suzaku team supplied instrumental tuned background file.  The source spectrum was corrected for instrument dead time.  We also included the contribution from the cosmic X-ray background, as suggested by the Suzaku team\footnote{See the Suzaku ABC Guide, Section 7.5.4 at \url{http://heasarc.gsfc.nasa.gov/docs/suzaku/analysis/abc/}}, as a cutoff power law of the form  \begin{math} {\rm CXB}(E) = 9.0 \times 10^{-9} \times (E/3 {\rm keV})^{-0.29}  \times \exp(-E/40 {\rm keV})~{\rm erg}~{\rm cm}^{-2}~{\rm s}^{-1}~{\rm sr}^{-1}~{\rm keV}^{-1}\end{math}.  The standard response file from the Suzaku CALDB was used and the spectrum was binned to a signal-to-noise ratio of 3\,$\sigma$.  The net exposure time for the PIN spectrum is 34\,ks with a count rate of 0.31\,counts\,s$^{-1}$.       

\subsection{The Swift BAT Spectrum}
To fully characterize the spectrum of NGC 6860, we also include an analysis of the Swift BAT spectrum generated from the first 22 months of the all-sky survey.  The BAT spectrum is an 8-channel spectrum in the 14--195\,keV band.  Both the spectrum and diagonal matrix are further described in the 22-month catalog paper \citep{2010ApJS..186..378T}.

\section{Light Curve Analysis}~\label{var}
In our previous analysis of variability in NGC 6860 \citep{2008ApJ...674..686W}, we found that the EPIC MOS spectra were variable over the $\approx 10$\,ks observation.  We also found that a comparison of this short XMM-Newton observation with the Swift XRT observation showed a change in the soft (0.5--2\,keV) flux with no change in the hard (2--10\,keV) flux.  The difference between these observations was about 3 months.  Therefore, the source is known to vary in the soft band both on short timescales and over months.

\begin{figure}
\begin{center}
\includegraphics[width=11cm]{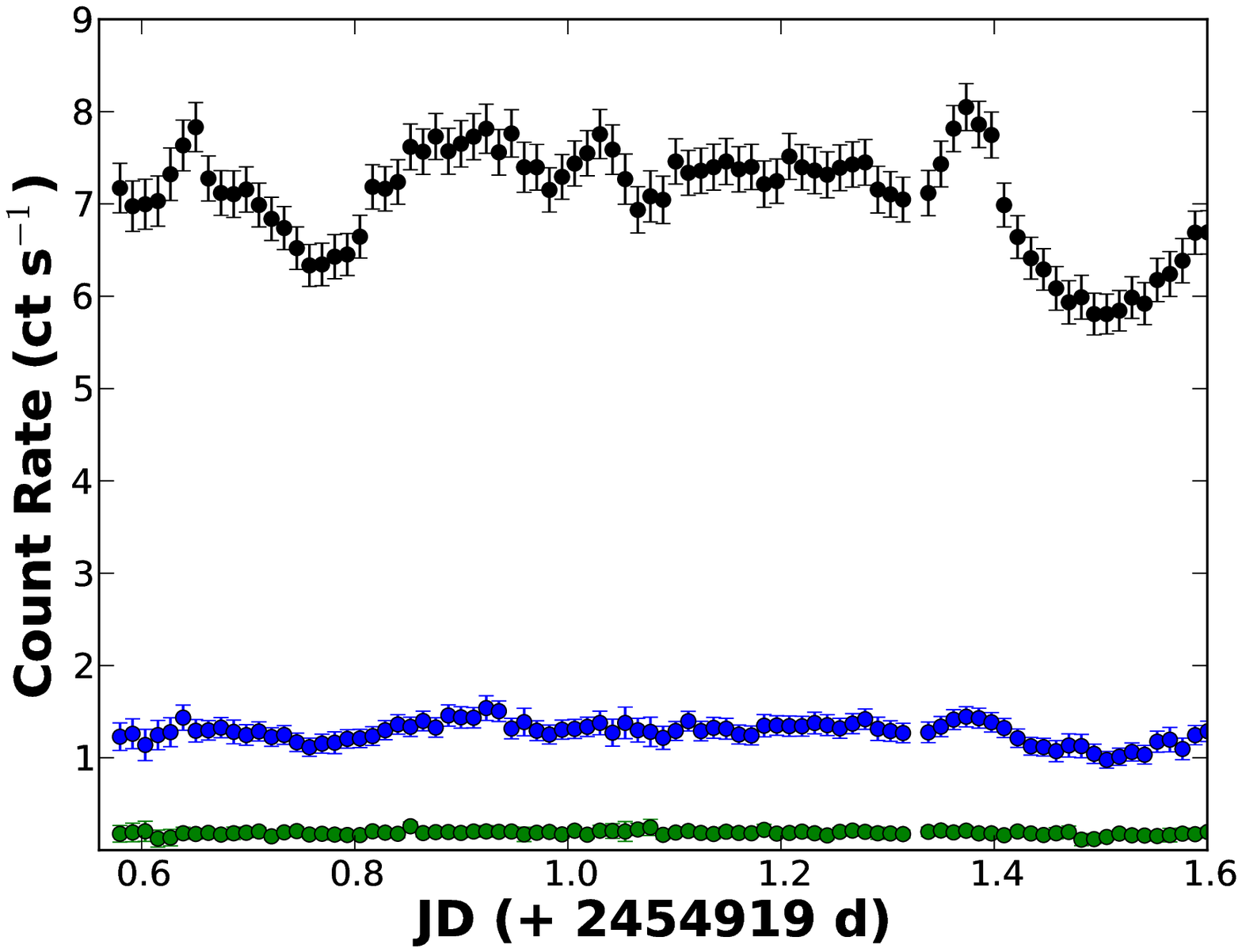}\\
\includegraphics[width=11cm]{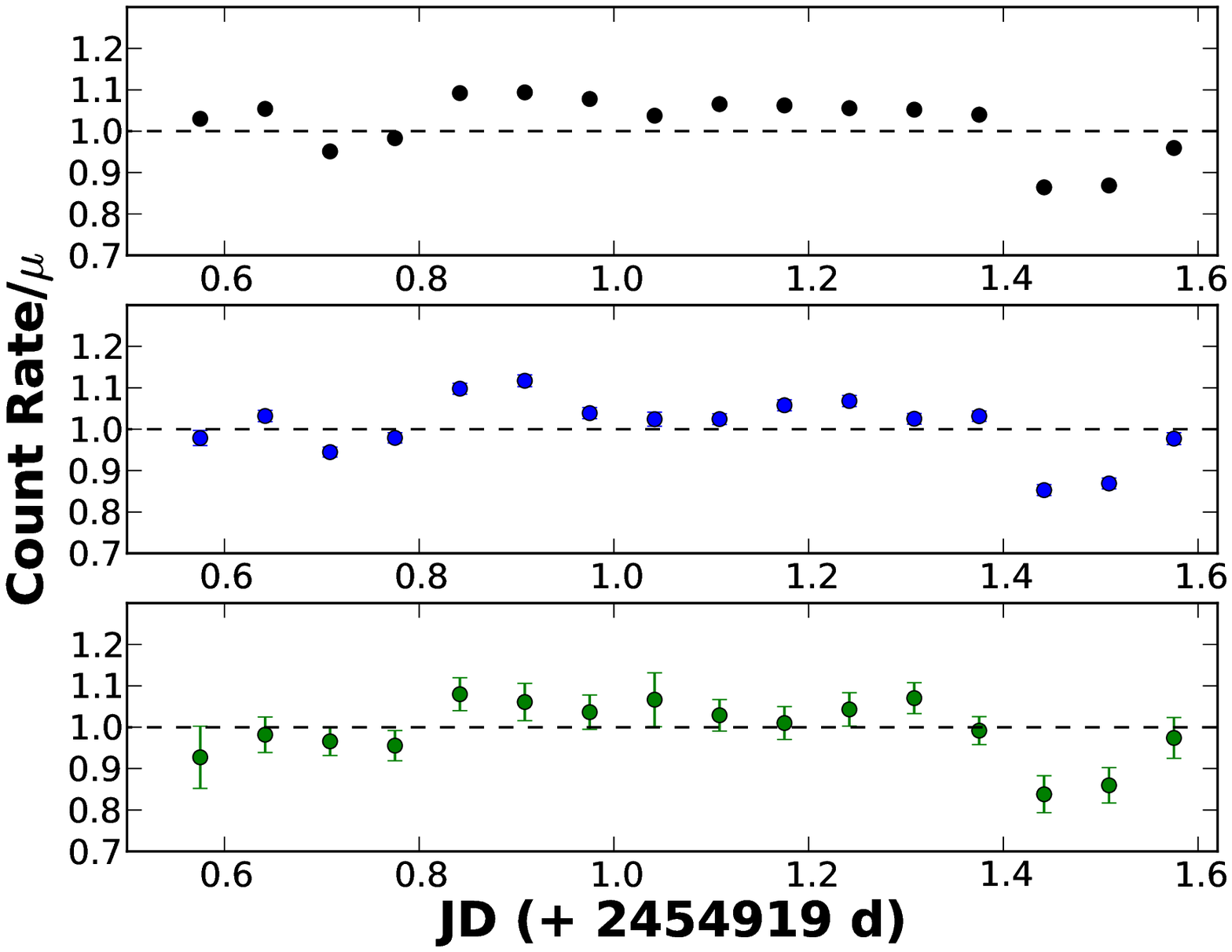}
\end{center}
\caption{The XMM-Newton pn background subtracted light curves are shown binned by 1024\,s (top) and 5760\,s (bottom, normalized by the average count rate in the specified energy band).  Light curves are shown for three selected energy bands: 0.1--3\,keV (black), 3--7\,keV (blue), and 7--10\,keV (green).  Clearly, the greatest amount of variability is seen in the soft band, which varies by about 2\,cts\,s$^{-1}$.  However, the normalized light curves (bottom) show that the relative amount of variability is the same in each of the bands.  Further, each of the normalized light curves vary on the same timescale.  We have excluded from the plot regions of high background flares.
\label{fig-pnlc}}
\end{figure}

To determine the nature of variability over the long XMM-Newton observation, we constructed light curves for the EPIC pn and MOS spectra of NGC 6860.  We constructed light curves in three separate bands: soft (0.1--3\,keV), medium (3--7\,keV), and hard (7--10\,keV).  In this way, we can associate the amount of variability with features prominent in each of these bands -- namely, a soft excess and/or warm absorbers in the soft band, the region surrounding the Fe K$\alpha$ region for the medium band, and the hard continuum in the hard band.  In Figure~\ref{fig-pnlc}, we plot the background subtracted pn light curves, binned on a short 1024\,s timescale and the orbital timescale (5760\,s, showing the normalized light curves).  The mos light curves, which are not shown, exhibit the same trends seen in the pn.  

The light curves show that the soft band dominates the total count rate and that this band varies by count rates of as much as 2\,counts\,s$^{-1}$ over the observation.  The mean count rates in each of the bands correspond to 6.9\,counts\,s$^{-1}$ (soft), 1.3\,counts\,s$^{-1}$ (medium), and 0.2\,counts\,s$^{-1}$ (hard).  When we look at the normalized light curves, we find that the variability in the soft band correlates well with the variability in both the medium and hard bands.  

\begin{figure}
\centering
\includegraphics[width=9cm]{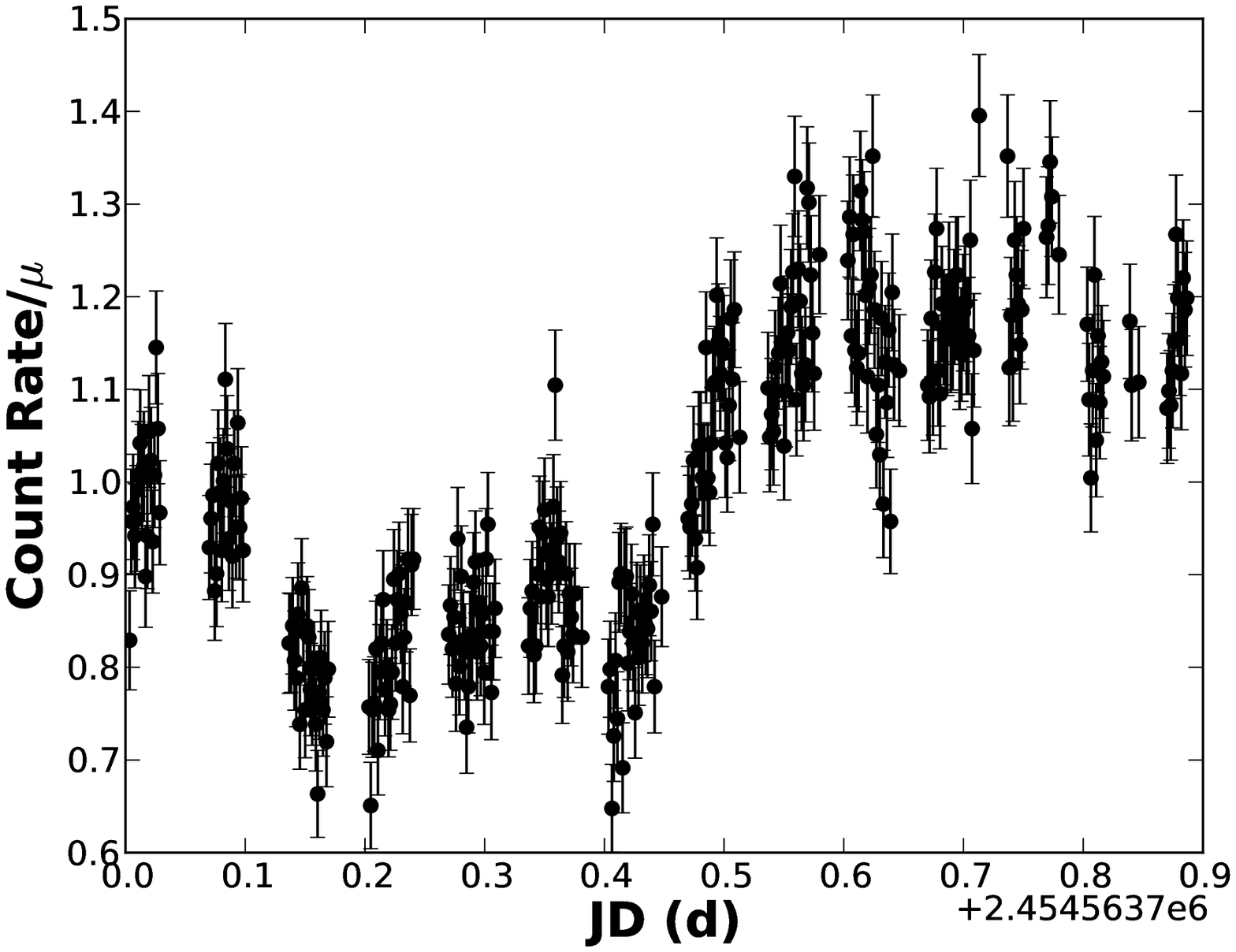}\\
\includegraphics[width=9cm]{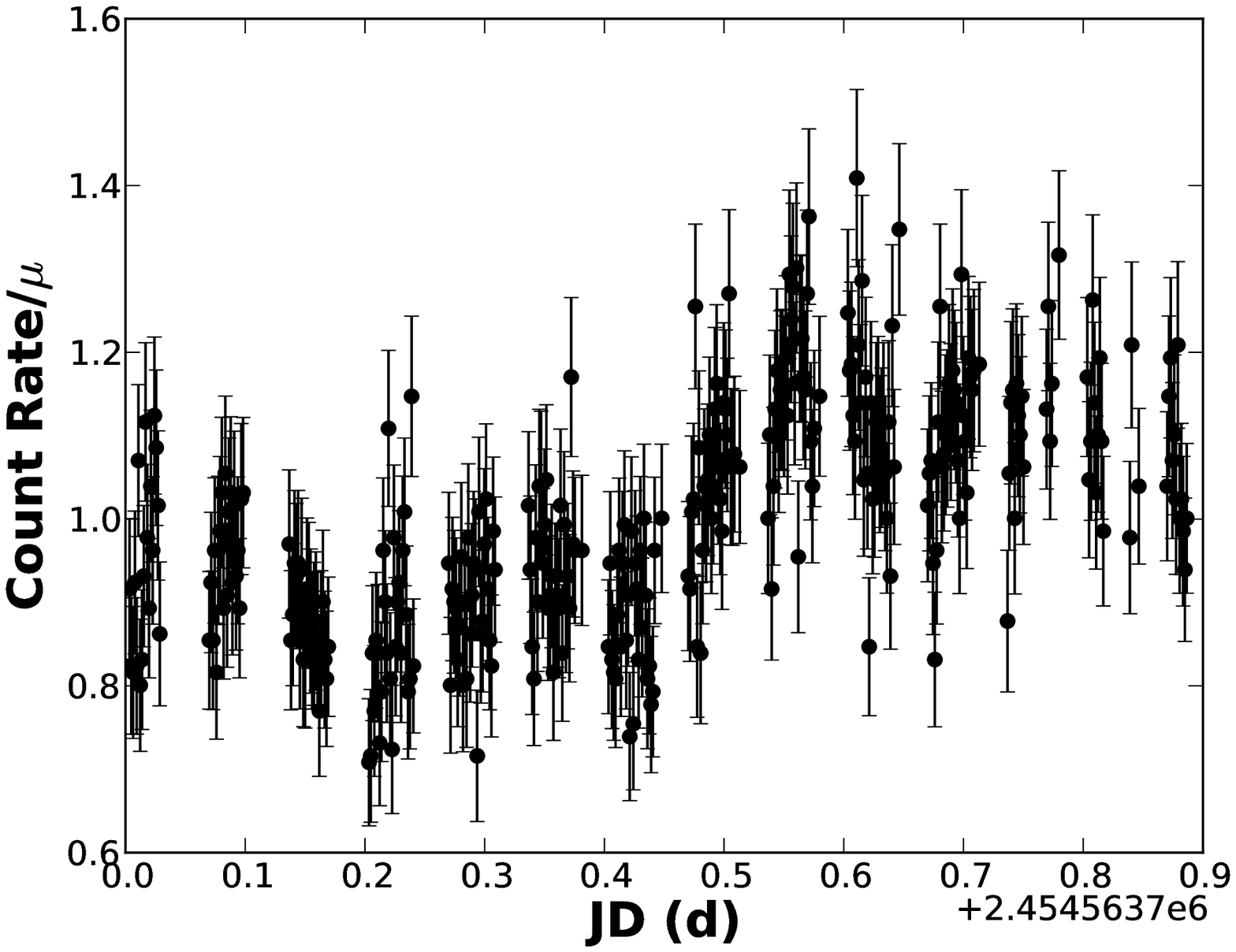}\\
\includegraphics[width=9cm]{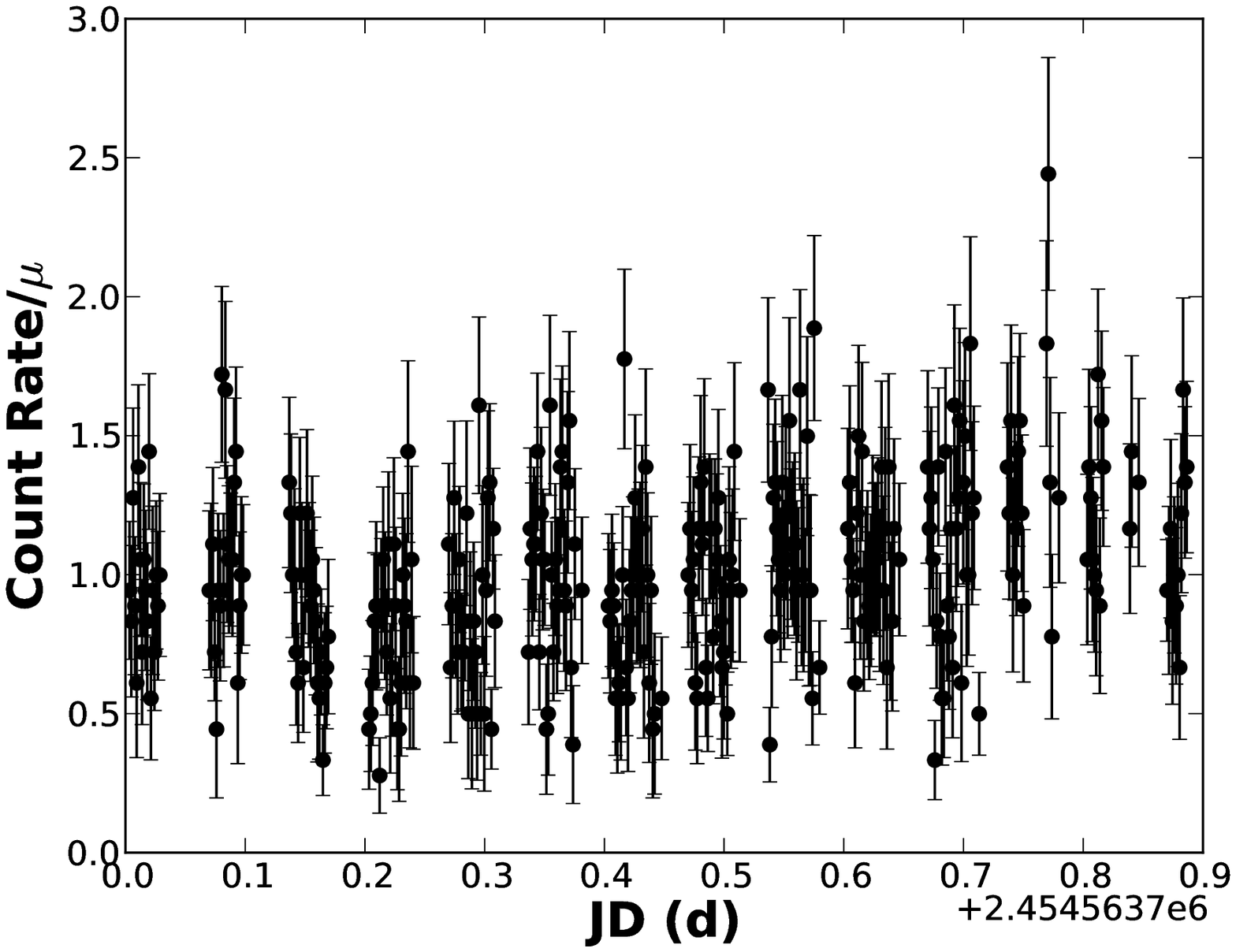}\\
\caption{The Suzaku XIS (XIS0+XIS1+XIS3) normalized background subtracted light curves are shown in the soft band (0.1--3\,keV, top), medium band (3--7\,keV, middle), and hard band (7--12\,keV, bottom).  Each of the light curves are binned by 128\,s.  As in the XMM-Newton observation, we find that variability in each of the bands is well-correlated.
\label{fig-suzlc}}
\end{figure}

In a similar way as for the XMM-Newton light curves, we constructed light curves in a soft, medium, and hard band for the Suzaku XIS detectors.  Since the response for the XIS detectors extends to higher energies, we used the energy range of 7--12\,keV for our XIS hard band.  We created a combined (XIS0 + XIS1 + XIS3) background subtracted light curve for the Suzaku observation, binned by 128\,s.  As for the XMM-Newton observation, we found that the low energy band had the highest average count rate ($< {\rm Ct\,Rate}> = 0.83$\,cts\,s$^{-1}$).  Background subtracted mean count rates in the medium and hard band correspond to 0.34\,cts\,s$^{-1}$ and 0.05\,cts\,s$^{-1}$, respectively.  In Figure~\ref{fig-suzlc}, we plot the normalized 128\,ks light curves.  As for the XMM-Newton observations, we find that the variability in the soft, medium, and hard bands is well-correlated, particularly for the soft and medium bands.

\section{Spectral Analysis}~\label{spectra}
Given the complex nature of the spectra of NGC 6860, we began our spectral fits by analyzing the spectra in the soft and hard energy bands separately.  We first determine the parameters of the hard X-ray continuum and the region surrounding the Fe K$\alpha$ emission in Section~\ref{hard}.  This analysis includes spectra from Suzaku and XMM-Newton, focusing on the spectral region from 3--50\,keV.  To determine the properties of the soft band, the region corresponding to the highest count rates for NGC 6860 and the greatest complexity, we examine the XMM-Newton and Suzaku CCD spectra as well as the XMM-Newton RGS spectra in Section~\ref{soft}.  We then fit the broadband continuum, including the Swift BAT spectra, in Section~\ref{broad}.  We use the best broad band fit to constrain the continuum in the RGS data, to better determine the properties of the warm absorbers in Section~\ref{better}.  All of the spectral fits described were done using XSPEC version 12.5.1 \citep{ADASS_96_A}.  The quoted errors correspond to a change in $\chi^2 = 2.71$ or the 90\% confidence level for an additional free parameter.

\subsection{The Hard X-ray Continuum and the Fe K$\alpha$ Emission Region}~\label{hard}
In the following sub-sections, we describe spectral fits to the hard X-ray emission ($> 3$\,keV) for the Suzaku and XMM-Newton observations.  We include absorbed power law models, as well as fits with a cutoff energy/reflection component.  Also, we fit for additional lines/absorption edges in the region around the 6.4\,keV \ion{Fe}{1} K$\alpha$ emission.

\subsubsection{The Suzaku spectra}
We conducted combined fits of the Suzaku XIS (XIS1 and XIS0+XIS3) and PIN spectra, constraining the PIN data such that its spectrum is fixed at a constant level of 1.16 of the XIS1 spectrum, as indicated in the Suzaku Data Reduction Guide\footnote{\url{http://heasarc.gsfc.nasa.gov/docs/suzaku/analysis/abc/}}.
We find that a simple absorbed power law model provides a good fit to the data (reduced $\chi^2/{\rm dof} = 1.09/1085$).    Fixing the Galactic absorption to the \citet{1990ARAA..28..215D} measured value of $4.19 \times 10^{20}$\,cm$^{-2}$ with a multiplicative {\tt tbabs} model
\citep{2000ApJ...542..914W}, we find that the neutral hydrogen column density of the source is best fit as $4.8^{+4.2}_{-4.3} \times 10^{21}$\,cm$^{-2}$ with the multiplicative {\tt ztbabs} model \citep{2000ApJ...542..914W}.  The power law component measured is typical of the Swift BAT-detectected AGN 
\citep{2009ApJ...690.1322W}, with $\Gamma = 1.70 \pm 0.05$ (from the {\tt zpowerlw} model).  Residuals to the model indicate the presence of an Fe K$\alpha$ emission line.

Adding a Gaussian to account for the Fe K$\alpha$ emission with the {\tt zgauss} model, we find that the line is significant with an improvement in the fit by $\Delta \chi^2 = -48$.  The measured parameters for this line are $E = 6.31 \pm 0.05$\,keV, $\sigma = 0.15^{+0.06}_{-0.04}$\,keV, $I = 2.6^{+0.6}_{-0.8} \times 10^{-5}$\,photons\,cm$^{-2}$\,s$^{-1}$, and $EW = 102^{+24}_{-32}$\,eV.
The width of this line indicates an origin interior to the broad line region (with AGN H$\beta$ FWHM values typically ranging from 3000 -- 10,000\,km\,s$^{-1}$ and measured as $5920 \pm 600$\,km\,s$^{-1}$ for NGC 6860 \citep{2006AA...459...55B}), with a corresponding velocity of $\approx 17000$\,km\,s$^{-1}$.  This broad width is statistically a better fit than a narrow line fit, with a $\Delta\chi^2 = -7.6$ over a fixed width of $\sigma = 0.01$\,keV.

The addition of a 7.11\,keV Fe edge (over the edge built into the {\tt ztbabs} model), is also significant, with $\Delta\chi^2 = -6.1$.   The energy of the edge is consistent with cold Fe
\citep{1993ADNDT..54..181H}, with $E = 7.11^{+0.19}_{-0.18}$\,keV.  The measured optical depth corresponds to $\tau = 0.08 \pm 0.05$.  The Fe edge can be the result of either reflection of the direct emission, a higher than solar iron abundance, or a higher column density of gas than measured from the neutral absorption model.  Computing the \ion{Fe}{17} cross section at 7.11\,keV following \citet{1996ApJ...465..487V}, we find a cross section of $\sigma(E) = 4.5 \times 10^{-3}$\,Mb.  From the estimated cross section and our measured optical depth, we find a column density of iron that is $1.80 \times 10^{19}$\,cm$^{-2}$ ($N_{\rm Fe} = \tau/\sigma(E)$).  Therefore, the estimated iron abundance is very high ($N_{\rm Fe}/N_{\rm H} \approx 139$\,(Fe/H)$_{\sun}$) if the edge is associated with the $4.8 \times 10^{21}$\,cm$^{-2}$ absorber measured in our fits.  If, however, the abundance of iron in NGC 6860 is the solar value, this implies that the Fe edge is instead the result of a much higher hydrogen column density of $\approx 7 \times 10^{23}$\,cm$^{-2}$, which does not affect the emission in the softer bands (i.e.~a partially covering near Compton-thick absorber).  Finally, while the residuals to the model suggest there may be additional emission/absorption features present, we found no statistical support for additional emission features found in the spectra of many AGNs (\ion{Fe}{25} K$\alpha$, \ion{Fe}{1} K$\beta$, \ion{Ni}{1} K$\alpha$).

In addition to a simple absorbed power law model, we tried to determine whether a high energy cutoff or reflection component are present in the data.  To constrain a possible high energy cutoff, we substituted the {\tt zpowerlw} model with a cutoff power law model.  However, we found no measurable improvement in the fit and could not constrain the energy of the cutoff.    

Additionally, we tested for the presence of a reflection component, which could alternatively account for the presence of the 7.11\,keV Fe edge.  Using the {\tt pexrav} model for reflection off of cold neutral material, we tied the normalization and power law index to be the same values between the direct power law component and the reflection component.  Statistically, the reflection model does not significantly change the fit, improving $\chi^2$ by only 4.4, while losing two degrees of freedom.  Using this model, we find neither the folding energy ($E_{fold} = 30^{+89}_{-19}$\,keV) nor the reflection parameter (which can range from 0--1 with no change in $\chi^2$) are well constrained.  These fits, however, do not include the higher energy BAT data, which are used to place tighter constraints on reflection and the energy cutoff in Section~\ref{broad}.

\subsubsection{The XMM-Newton EPIC spectra}
\begin{figure}
\centering
\includegraphics[width=15cm]{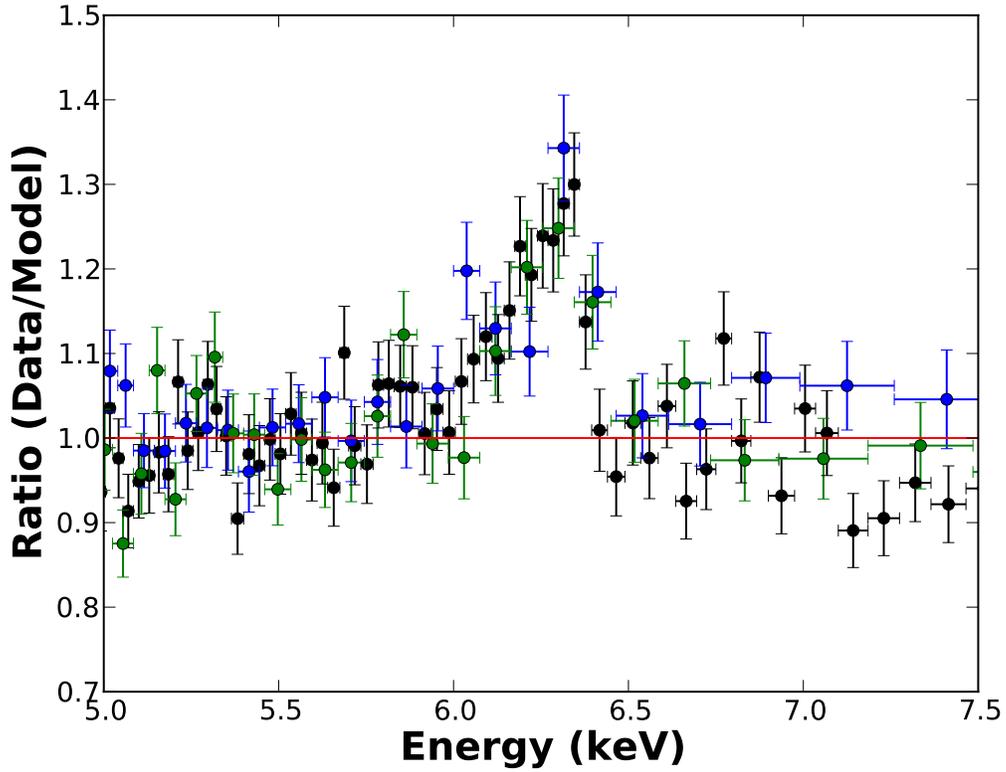}
\caption{The ratio of the data/model for the XMM-Newton EPIC spectra, fit by a simple absorbed power law from 3--10\,keV, is shown for an absorbed power law model.    The pn (black), MOS1 (blue), and MOS2 (green) data are shown specifically around the Fe K$\alpha$ emission line region.  Clearly, the Fe K$\alpha$ emission line is present.  A narrow emission line at 5.95\,keV is also significant ($\Delta\chi^2 = -15$).  
\label{fig-res}}
\end{figure}

For spectral fits of the hard continuum in the EPIC spectra, we simultaneously fit the pn, mos1, and mos2 spectra in the 3--10\,keV band, using a constant value to allow for variations in the flux levels between CCDs.  We find a good fit (reduced $\chi^2/{\rm dof} = 1.17/2336$) with a simple absorbed power law model.  The parameters of the model indicate a column density (above the Galactic value) and power law index similar to those measured in the Suzaku observation: $N_{\rm H} = 5.8^{+2.7}_{-2.6} \times 10^{21}$\,cm$^{-2}$ and $\Gamma = 1.69\pm 0.04$.  The residuals to this model, shown in Figure~\ref{fig-res}, clearly show an Fe K$\alpha$ emission line. Additionally, residuals suggest an emission feature red-ward of the neutral Fe K$\alpha$ emission and possibly an 7.11\,keV Fe edge (in the pn spectrum).  Fitting an Fe edge at 7.11\,keV, however, results in no significant change in $\chi^2$ of the model.  We find an upper limit on the optical depth of the edge at $\tau = 0.07$, which is in agreement with the Suzaku results.

The addition of a Fe K$\alpha$ emission line is clearly significant, resulting in $\Delta\chi^2 = -190.6$.  The measured line parameters are again similar to those found in the Suzaku observation.  We find $E = 6.35\pm0.02$\,keV, $\sigma = 0.11^{+0.03}_{-0.02}$\,keV, $I = 2.2^{+0.4}_{-0.3} \times 10^{-5}$\,photons\,cm$^{-2}$\,s$^{-1}$, and $EW = 85^{+14}_{-12}$\,eV.  The line is clearly broad, since there is a $\Delta\chi^2 = 15$ between the fitted broad $\sigma$ and a fixed narrow line ($\sigma = 0.01$\,keV).  The best-fit energy, width, and equivalent width are consistent between the XMM-Newton and Suzaku observations.  Since we would expect an Fe K$\beta$ emission line to also be present if the Fe K$\alpha$ emission is produced in neutral material, we added this feature to our spectral fits.  We fixed the centroid energy to 7.056\,keV, the width to 0.01\,keV, and the intensity to be 13\% of the Fe K$\alpha$ line.  There is virtually no change in $\chi^2$ ($\Delta\chi^2 = -0.01$) and no change in the parameters of this fit upon adding the Fe K$\beta$ line.

Adding a narrow Gaussian ($\sigma = 0.01$\,keV) to fit the additional emission line seen in the residuals, we find that the line is significant, improving $\chi^2$ by 15.  The fitted parameters of this emission feature are $E = 5.95^{+0.03}_{-0.06}$\,keV, $I = 5.1^{+2.7}_{-1.3} \times 10^{-6}$\,photons\,cm$^{-2}$\,s$^{-1}$, and $EW = 17^{+4}_{-9}$\,eV.  While the line is weak, it is also significant.  The energy of the line is unusual, but corresponds to lines seen in the spectra of other AGN sources.  For instance, as pointed out in \citet{2009ApJ...691..922M}, high resolution spectroscopy of 16 Seyferts have shown narrow features interpreted as blueshifted or redshifted Fe K features from hot spots above the accretion disk \citep{2008MNRAS.390..421V}.  The line we have detected has roughly the same energy, width, and strength as the 6.0\,keV feature detected in NGC 3227 by \citet{2009ApJ...691..922M}.  

Comparing the flux from the best fit absorbed power law + Fe K$\alpha$ emission model with XMM-Newton to the Suzaku spectra, we find no change in flux.  The XMM-Newton observation has a 3--10\,keV observed flux of $1.86 \times 10^{-11}$\,ergs\,s$^{-1}$\,cm$^{-2}$ in the pn.  The XIS spectra have an average flux of $1.84 \times 10^{-11}$\,ergs\,s$^{-1}$\,cm$^{-2}$.  Therefore, while the XMM-Newton observations were taken a year after the Suzaku observations, there is virtually no change in the hard X-ray spectra.

\subsection{The Soft X-ray Spectra}~\label{soft}
In the following sub-sections, we detail simple model fits to the soft spectra from the Suzaku and XMM-Newton observations.  The base model is an absorbed power law, at the redshift of the source.  To account for the additional features in the soft spectra, we use a combination of absorption edge models and Gaussians to fit absorption and emission features.

\subsubsection{The Suzaku XIS spectra}
\begin{figure}
\centering
\includegraphics[width=15cm]{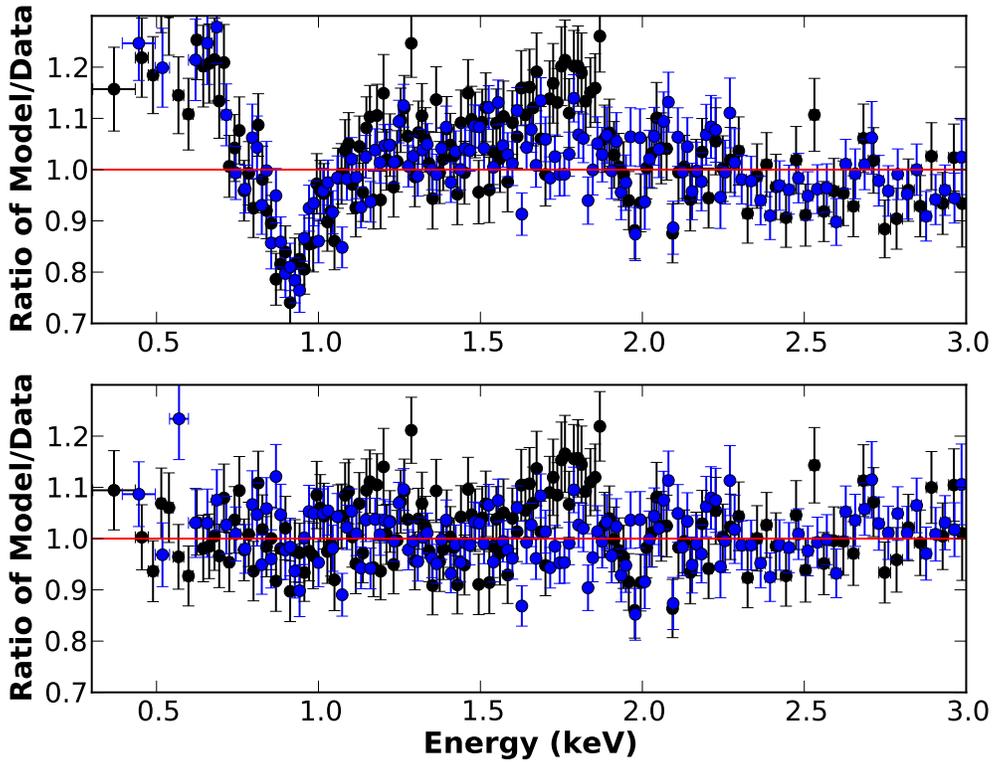}
\caption{Ratios of the data to model are plotted for spectral fits to the Suzaku observations in the 0.3-3\,keV band.  In the top plot, the ratio of the data/model for a simple absorbed power law model is shown.  The fit is greatly improved with the addition of \ion{O}{7} and \ion{O}{8} edges, as shown in the bottom plot.  The Suzaku XIS0+XIS3 spectrum is shown in black and the XIS1 spectrum is shown in blue.  Also present in the spectrum is an absorption edge at 1.91\,keV, which is likely an instrumental feature associated with the Si K edge.
\label{fig-softsuzaku}}
\end{figure}

Fitting the soft XIS spectra simultaneously in the 0.3-3\,keV band, we find that a simple absorbed power law model is not a good fit to the data with a reduced $\chi^2/{\rm dof} = 1.64/883$.  Comparing the fitted parameters of this model to the hard continuum, we find that both the neutral hydrogen column density ($N_{\rm H} = 2.9\pm0.9 \times 10^{20}$\,cm$^{-2}$)  and power law index ($\Gamma = 1.36\pm0.03$) are lower in the soft band.  As shown in Figure~\ref{fig-softsuzaku}, the simple power law model clearly can not account for the soft spectrum.

We find that the addition of warm absorption signatures, through \ion{O}{7} and \ion{O}{8} edges, to the model greatly improves the fit.  Figure~\ref{fig-softsuzaku} shows the improvement in the fit for the ratio of data/model.  With these edges added to the model, we find $\Delta\chi^2 = -514.7$ and a reduced $\chi^2/{\rm dof} = 1.07/879$.  Additionally, while the column density of the intrinsic absorption is still low ($N_{\rm H} = 7.9^{+1.0}_{-0.9} \times 10^{20}$\,cm$^{-2}$), the power law index from this warm absorber model is consistent with the hard power law continuum ($\Gamma = 1.74\pm 0.04$).  The best-fit parameters for the oxygen edges correspond to $E = 0.73 \pm 0.01$, $\tau = 0.34 \pm 0.04$ (\ion{O}{7}) and $E = 0.87 \pm 0.01$, $\tau = 0.35^{+0.03}_{-0.04}$ (\ion{O}{8}).  In addition to the \ion{O}{7} and \ion{O}{8} edges, the residuals to the model suggest another edge feature near 2\,keV.  Adding an edge improves the fit by $\Delta\chi^2 = 10$.  This edge, which has best-fit parameters of $E = 1.91^{+0.04}_{-0.05}$\,keV and $\tau = 0.055^{+0.030}_{-0.027}$, is likely associated with errors in the detector calibration of the Si K edge.  For our final best-fit model, the estimated observed flux in the 0.3--3\,keV band is $1.34 \times 10^{-11}$\,ergs\,s$^{-1}$\,cm$^{-2}$.

\subsubsection{The XMM-Newton EPIC spectra}
\begin{figure}
\centering
\includegraphics[width=15cm]{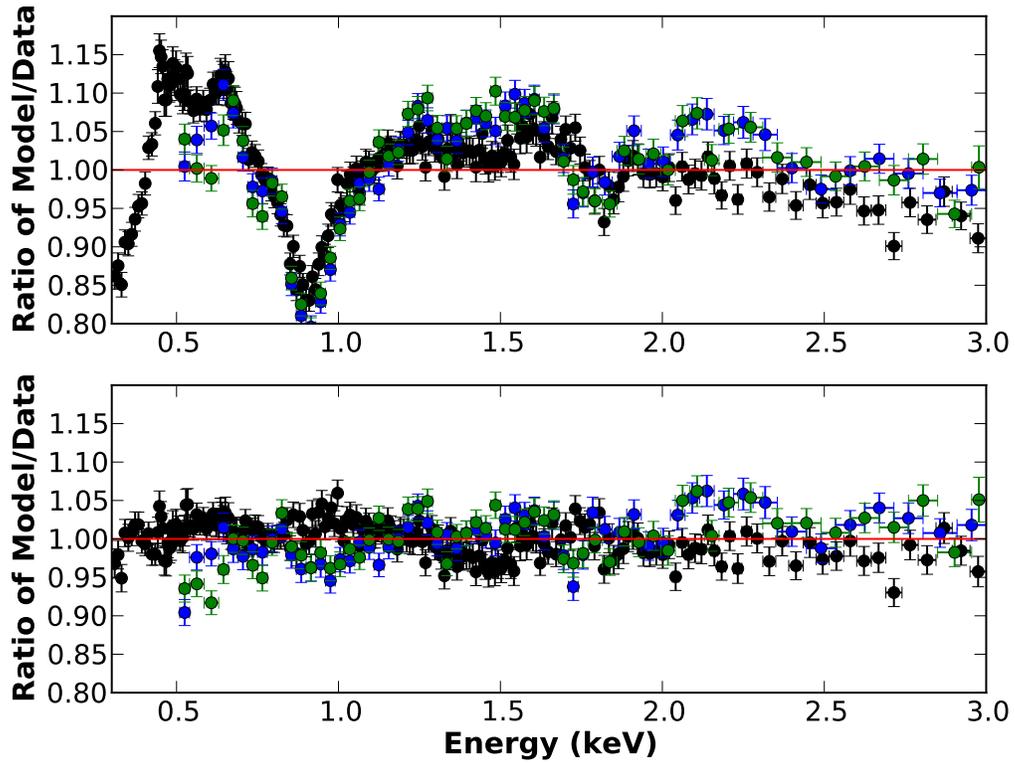}
\caption{Ratios of the data to model are plotted for spectral fits to the XMM-Newton EPIC observations in the 0.3-3\,keV band (rebinned by 50 for illustrative purposes).  In the top plot, the ratio of data/model points is shown for a simple absorbed power law model.  The fit is greatly improved with the addition of \ion{O}{7} and \ion{O}{8} edges, in addition to additional lines described in Table~\ref{tbl-softepic}, as shown in the bottom plot.
\label{fig-softxmm}}
\end{figure}

Given the much higher signal-to-noise in the XMM-Newton EPIC spectra, a simple absorbed power law model does not adequately describe the 0.3--3\,keV EPIC spectra, with a reduced $\chi^2/{\rm dof} = 7.71/898$.  In Figure~\ref{fig-softxmm}, we plot the ratio of the data/model for the simple absorbed power law model.  As for the Suzaku spectrum, the best fit requires a lower neutral hydrogen column density ($N_{\rm H} = 2.4 \times 10^{20}$\,cm$^{-2}$) and a flatter power law continuum ($\Gamma = 1.46$).  However, as shown in the figure, there are more features than can be accounted for with the addition of \ion{O}{7} and \ion{O}{8} edges.

While adding the \ion{O}{7} and \ion{O}{8} edges improves the fit significantly, with $\Delta\chi^2 = -4458.0$, the fit is still not acceptable (reduced $\chi^2/{\rm dof} = 2.75/894$, see Figure~\ref{fig-softxmm}).  With this higher signal-to-noise data, a better interpretation for the residuals between 0.7--1.0\,keV, from the absorbed power law model, is that this feature is from a broad unresolved transition array (UTA) from Fe M-shell ions.  Such features are detected in the {\it XMM-Newton} and Chandra spectra of a number of Seyfert 1 sources (e.g. NGC 3227, NGC 3783, NGC 4051, Mrk 509). 

We find that the addition of Gaussians/edges to the data greatly improves the fit.  However, we note that due to inconsistencies in the pn and MOS calibration below 0.5\,keV, we used only the pn data to determine properties of lines detected in this low energy region.  In Table~\ref{tbl-softepic}, we specify the parameters of lines/edges where $|\Delta\chi^2| > 15$.  These features include \ion{N}{6}, \ion{Ne}{9}, and \ion{Si}{13} absorption lines.  We found the addition of another edge model, at 1.7\,keV, improved the fit greatly ($\Delta\chi^2 = -26$).  The most probable identification for this feature is the \ion{Mg}{11} edge with a rest energy of 1.761\,keV, which is not near any known energies corresponding to calibration issues for the EPIC pn.  The improved fit results in a reduced $\chi^2/{\rm dof} = 1.45/855$.

From 0.45 to 0.5\,keV, the residuals showed complex absorption/emission that could not be resolved in the pn.  We found that an emission line at 0.45\,keV greatly improved the fit.  Given the uncertainties in the pn calibration below 0.5\,keV (see the EPIC calibration document\footnote{\url{http://xmm2.esac.esa.int/docs/documents/CAL-TN-0018.pdf}}), we can not uniquely identify this feature at 0.45\,keV.  However, we note that this is possibly \ion{N}{7}.

The associated outflow velocities determined from the fitted energies of the edges and lines are difficult to interpret.  We find that the velocities of the \ion{O}{7} and \ion{O}{8} edges are consistent with each other, implying red-shifted velocities of approximately 8000--9000\,km\,s$^{-1}$.  The estimated velocity of the \ion{Mg}{11} edge is consistent with the oxygen edges. However, the energy of the \ion{Ne}{9} emission line suggests a blue-shifted velocity of about 5000\,km\,s$^{-1}$ while the remaining lines, are consistent with no velocity shift.  The inconsistent velocities are likely due to a combination of poor energy resolution in the EPIC spectra and confusion of multiple features.  The 20--30\,eV shifts in the fits of the spectral features seen in the pn data are probably due to instrumental effects, since they are not seen in the Suzaku data.  Shifts this large, while rare, have been seen in pn, particularly during episodes of high background.    

The best fit neutral column density is much lower than that found from fits to the hard spectrum (we find N$_{\rm H} = 5.93^{+0.09}_{-0.07} \times 10^{20}$\,cm$^{-2}$).  However, we find that the best-fit soft band power law index is consistent with the hard band value (we find $\Gamma = 1.697^{+0.004}_{-0.003}$).  The estimated flux in the 0.3--3\,keV band is $1.31 \times 10^{-11}$\,ergs\,s$^{-1}$\,cm$^{-2}$, which is also consistent with the soft flux in the Suzaku observation.

\subsubsection{The XMM-Newton RGS spectra}
As a first fit to the RGS spectra, we fit an absorbed power law model ({\tt zpow}), with absorption fixed at the Galactic value.   Unlike with the EPIC CCD data, we find a good fit ($\chi^2/{\rm dof} = 2865.9/1960$) with a slope, $\Gamma = 1.50\pm0.02$.  The addition of an intrinsic neutral absorber does not improve the fit.  The addition of a blackbody component improves the fit significantly, with $\Delta\chi^2 = -105.3$.  Adding edge models to account for the \ion{O}{7} and \ion{O}{8} absorption edges further improves the fit with $\Delta\chi^2 = -129.3$.  The best-fit parameters are: \ion{O}{7} edge energy $= 0.705^{+0.002}_{-0.004}$\,keV with optical depth $\tau = 0.15^{+0.06}_{-0.04}$, \ion{O}{8} edge energy $= 0.850^{+0.002}_{-0.002}$\,keV with optical depth $\tau = 0.30^{+0.03}_{-0.04}$, and kT $= 0.30^{+0.04}_{-0.02}$\,keV with normalization of 1.9$^{+0.8}_{-0.7} \times 10^{-5}$\,photons\,cm$^{-2}$\,s$^{-1}$ at 1\,keV for the blackbody component.  The flux in the 0.3--2.5\,keV band is $1.22 \times 10^{-11}$\,ergs\,cm$^{-2}$\,s$^{-1}$.

\begin{figure}
\centering
\includegraphics[width=18cm]{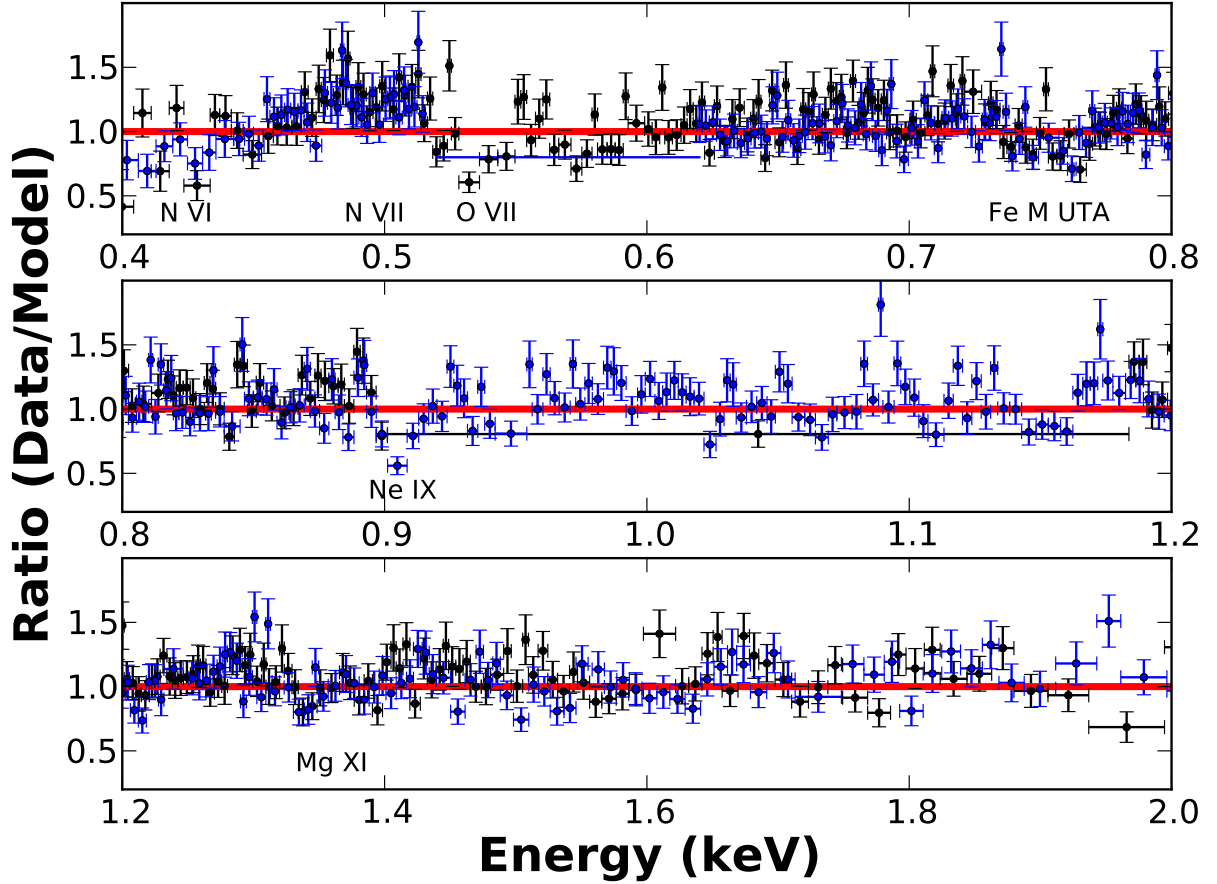}
\caption{The ratio of the data/model for the power law + blackbody model, along with the \ion{O}{7} and \ion{O}{8} edges, is plotted for the RGS spectra (re-binned to a signal-to-noise of 7 for illustrative purposes).  The observed energy is plotted, while lines and features are labeled in the AGN rest-frame energy.  The value where the data and model match (1.0) is plotted with a thick line.  RGS1 is shown in black and RGS2 is shown in blue.  Regions where the RGS1 or RGS2 have no data, are represented by a solid line.  Emission/absorption features that were added to improve the spectrum (i.e. \ion{Ne}{9} near 0.9\,keV, see Table~\ref{tbl-rgs}) are indicated.  Additionally, there appears to be an Fe M-shell UTA feature in the spectra between 0.7 and 0.8\,keV.  \ion{Mg}{11} also appears to be present in absorption, though fitting a Gaussian at this energy does not significantly improve $\chi^2$.
\label{fig-rgs}}
\end{figure}

Following this simple power law + blackbody fit (along with the \ion{O}{7} and \ion{O}{8} edges), we determined the significance of additional absorption and emission features seen in the residuals to the model.  We significantly detect (i.e. $|\Delta\chi^2| > 10$ when a Gaussian is added) \ion{N}{6}, \ion{O}{7}, and \ion{Ne}{9} in absorption and \ion{N}{7} in emission.  Details of these Gaussian fits, including the energy and intensity of the lines, are shown in Table~\ref{tbl-rgs}.  For all of these fits, we fixed the width of the lines to 0.01\,keV.  An additional absorption line of \ion{Mg}{11} appears in the residuals, but is not as significant in $\Delta\chi^2$.  We show the ratio of the data/model for the RGS spectra in Figure~\ref{fig-rgs}.  In addition to the features mentioned, there is evidence of an Fe M-shell UTA between 0.7 and 0.8\,keV.  While residuals to the model suggest that a \ion{Mg}{11} edge may be present in the RGS spectrum, adding the feature does not statistically improve the fit.

The combination of lower signal-to-noise in the RGS spectra and worse energy resolution in the EPIC spectra make it difficult to determine why some lines are only detected significantly in either the RGS or EPIC spectra.   However, with the exception of the \ion{Si}{13} line, the discrepancy in detection of lines is at the low energy range of the bandpass, from about 400-550\,eV.  In this region, there are uncertainties in the pn calibration.  Additionally, there are many features which are not easily resolved even with the higher resolution in the RGS spectra.  Therefore, we can not fully determine the emission and absorption properties of lines in this region without a higher signal-to-noise spectrum with the energy resolution of a grating spectrometer. The lack of a detection of \ion{Si}{13} absorption in the RGS could either be the result of the lower signal-to-noise of the RGS (the intensity of the line may be too low to detect significantly in the RGS spectra) or this line may be due to calibration uncertainties associated with the Si K edge in the EPIC detectors.  One clear conclusion we can draw from the comparison of the EPIC and RGS data is that the most significant features are the \ion{O}{7} and \ion{O}{8} edges and the \ion{Ne}{9} absorption line.

Unlike our analysis of the soft EPIC spectra, we find that the implied velocities from the oxygen edge energies do not agree with each other.  Assuming the lines are resonant lines, the energies of the \ion{N}{6} and \ion{O}{7} lines are consistent with each other, as well as the \ion{O}{8} edge, with blue-shifted velocities of about 7000\,km\,s$^{-1}$, similar to the velocities determined for the edges measured in the EPIC data.  The energy of the \ion{O}{7} edge suggests a velocity twice this value.  However, the energy of this edge is difficult to determine due to confusion with the Fe M-shell UTA (as seen in Figure~\ref{fig-rgs}).  Meanwhile, the energies of the \ion{Ne}{9} and \ion{N}{7} lines suggest low to no velocity shift.  These results suggest that there may be two different velocity components in the ionized absorber.  We further investigate the properties of the ionized absorbers with warm absorption model fits in Sections~\ref{sect-warmabs} and \ref{better}.

\subsubsection{Warm Absorber Model Fits}~\label{sect-warmabs}
Having fit the spectra of our sources with simple models for the absorption/emission lines in the soft spectra, we next investigate the properties of the warm absorbers through use of an analytic model.  The model {\tt warmabs} is an analytic version of the XSTAR table models \citep{1982ApJS...50..263K} available as additional models to XSPEC.  The model calculates spectra using stored level populations, which are scaled using the elemental abundances specified by the user.  The free parameters in the model include the column density of the absorber, the ionization parameter ($\log \xi$, where $\xi = {\rm L}/(n R^2)$ and $\xi$ has units of ergs\,s$^{-1}$), the redshift of the ionized gas, and the turbulent velocity of the gas (which affects the broadening of the spectral features).  For the {\tt warmabs} model, we assume a gas density of $10^{4}$\,cm$^{-3}$ and an illuminating power law with $\Gamma = 2$, as indicated in the population model {\it pops.fits.n04}.  We assume solar abundance for the elemental abundances, using the abundances of \citet{2000ApJ...542..914W} (in XSPEC, {\tt abund wilm}).  Additionally, we assume an outflow velocity of 0 km\,s$^{-1}$ for these initial fits, fixing the redshift of the warm absorber to the source redshift.  In Section~\ref{better}, we better constrain the continuum in the RGS spectra and re-fit the ionized absorbers allowing the velocity of the outflow to vary.

\begin{figure}
\centering
\includegraphics[width=18cm]{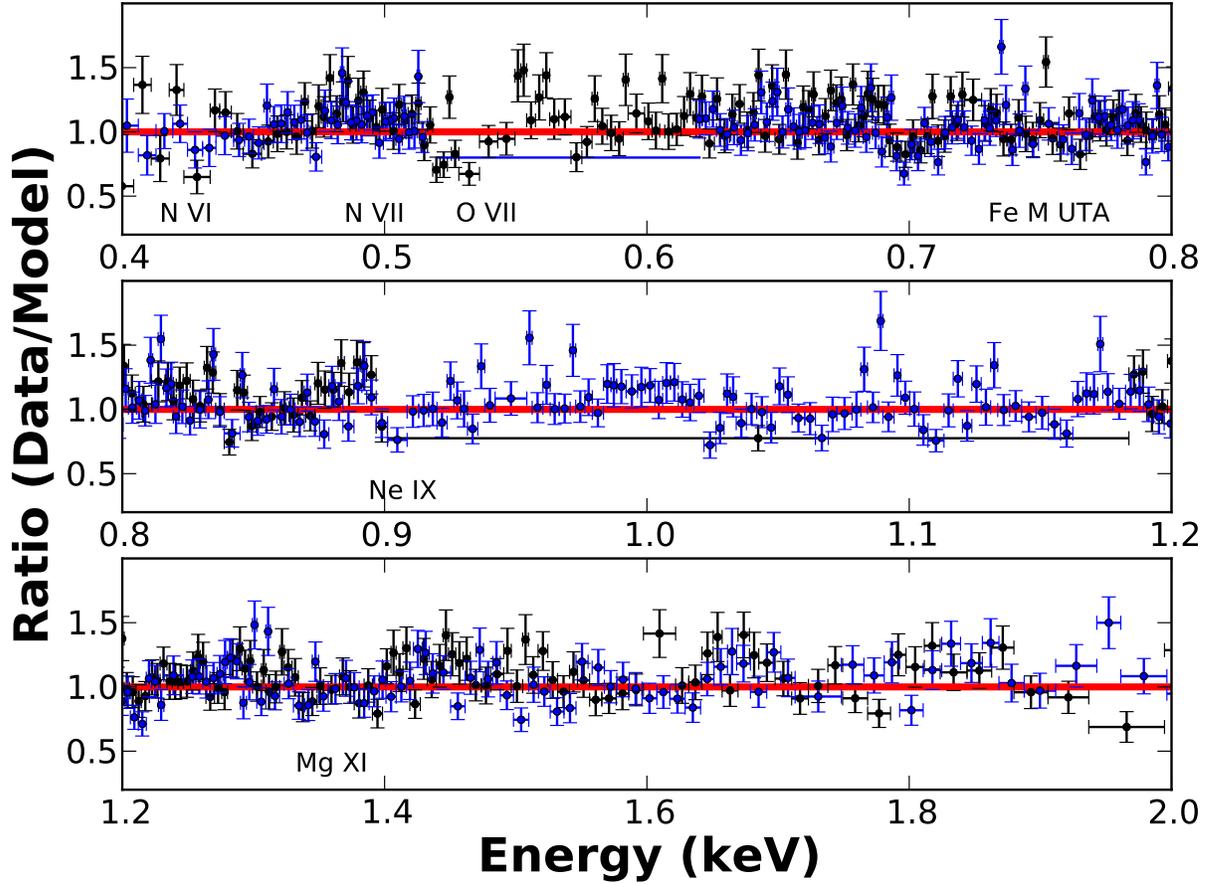}
\caption{The ratio of the data/model for the two component warm absorber model described in \S~\ref{sect-warmabs} are shown for the RGS spectra, re-binned to a signal-to-noise of 7. The observed energy is plotted, while lines and features are labeled in the AGN rest-frame energy. Overall, the two component warm absorber fit (along with a soft excess and power law continuum) is a good fit to the RGS spectra.  Some of the worst fits to the data are at the lowest energies, particularly below $\approx 0.7$\,keV.  The region near \ion{O}{7} show that this region is the least well fit by the model.  
Similar results are seen in the XMM-Newton EPIC pn spectra.
\label{fig-rgsresiduals}}
\end{figure}

We used this analytic warm absorption model to determine the best-fit parameters to the soft X-ray spectrum (0.3--3\,keV).  First, we fit each of the spectra individually (Suzaku XIS03+XIS1, XMM pn, and XMM RGS) with a base power law model, absorbed by the fixed Galactic absorption.  Since the RGS data required a black body fit, an indicator of the `soft excess', we added this as well as a component for neutral intrinsic absorption to the model fits.  The residuals to all of the spectra showed a poor fit, particularly to the Fe UTA feature.  Next, we added a single ionized absorber with the warm absorber model.  In all cases, the addition of a warm absorber model significantly improved the model fit and provided a good fit to the Fe UTA feature.  We then added a second warm absorber model, to determine if a second component is required.  For both the Suzaku XIS and XMM RGS spectra, the addition of a second component is statistically significant ($|\Delta\chi^2| \ga 50$, see Figure~\ref{fig-rgsresiduals} for the ratio of the data/model for the RGS spectra).  For the XMM pn spectrum, a second component is not required.  However, the residuals to the model for the warm absorption models show a poor fit below $\approx 0.7$\,keV.  The best fit parameters to the soft spectra with the warm absorption models are shown in Table~\ref{tbl-warmabs2}.

From our neutral + warm absorber model fits, we find that the neutral column does not change drastically between the Suzaku and XMM-Newton observations.  Additionally, the RGS and pn columns are fairly consistent (N$_{\rm H} \approx 10^{21}$\,cm$^{-2}$).  Comparing the XIS and pn observations, we find that the warm absorber (component 1 in the XIS) is consistent within the measured errors.  These values point to an ionized absorber with a column density of $\approx 10^{21}$\,cm$^{-2}$, an ionization parameter of $\log \xi \approx 2.0$, and a turbulent velocity from $\approx 25$--200\,km\,s$^{-1}$.  
The most notable differences between the Suzaku and pn spectra include the normalization on the black body component (which is roughly twice as high in the XIS spectra) and the slope of the power law (which is flatter in the pn observation).

There is clearly a mis-match between the best-fit parameters of the XMM pn and RGS spectra, which were taken simultaneously.  The inconsistencies are most apparent in both the blackbody (the RGS value is $\approx 1$\,keV, while the pn value is typical of the soft excess with $kT = 0.13$\,keV) and power law component (the RGS value is too steep, while the pn value is too flat).  With the warm absorber model, both the reduced $\chi^2$ (1.25 for RGS and 1.36 for pn) and residuals to the model show that the fits are not perfect representations of the spectra.  Further, the higher signal-to-noise in the pn and the better energy resolution in the RGS make it difficult to reconcile discrepancies between the fits.  

\subsection{Broad Band Spectral Fits}~\label{broad}
Since we found no significant difference in either flux or spectral shape between the Suzaku and XMM-Newton observations, we fit the Suzaku XIS and XMM-Newton pn data simultaneously with the Suzaku PIN and Swift BAT spectra to obtain the best broad-band (0.3--195\,keV) spectral fit.  A major problem of fitting non-simultaneous data, like the XMM-Newton pn and Swift BAT spectra, is constraining the proper normalization between the softer and hard energy spectra (see \citet{2008ApJ...674..686W} for a discussion of problems with joint XMM-Newton and Swift spectral fits).  This problem is mitigated by the addition of the Suzaku data, where the normalization between the PIN and XIS is known.  A second potential problem with using the Swift BAT spectra is that the BAT spectra are time averaged over a 22-month period.  Therefore, any changes in the spectral shape and/or flux will not be accounted for in the time averaged spectrum.  For NGC 6860, however, the Swift BAT 54-month light curve shows no significant variation (private communication, Gerry Skinner).

For the broadband spectral fits, we used a constant value ({\tt const}) to account for differences between the Suzaku XIS and the additional spectra.  This constant was set to 1.16 for the Suzaku PIN spectrum but allowed to vary for the XMM-Newton pn and Swift BAT spectra.  Our base model is a power law affected by the Galactic neutral column density.  To account for the warm absorber features, we added edges to account for the \ion{O}{7} and \ion{O}{8} absorption edges and Gaussians to account for the \ion{N}{7} and \ion{Ne}{9} absorption lines found in the XMM-Newton pn and RGS spectra.  A blackbody model was added to account for the soft excess.  Also, we added a Gaussian to account for the Fe K$\alpha$ emission near 6.4\,keV. 

The addition of an intrinsic absorption model (with the {\tt ztbabs} model) to the spectral fits greatly improves the fit with $\Delta\chi^2 = -3025.8$.  The fitted column density, N$_{\rm H} \approx 6 \times 10^{20}$\,cm$^{-2}$, is consistent with the neutral column derived from the warm absorber model fits to the soft spectra.  To determine if the absorber is partially or fully covering the spectrum, we tried substituting the fully covering {\tt ztbabs} model with the {\tt pcfabs} model.  Previously, in 
\citet{2009ApJ...690.1322W}, we found that the partial covering model was successfully applied to the majority of complex spectral sources in the Swift BAT AGN catalog.  The partial covering absorption model ({\tt pcfabs}) has two parameters, the covering fraction ($f$) and the neutral hydrogen column density (N$_{\rm H}$), such that: \begin{math}
M(E) = f \exp(-{\rm N}_{\rm H} \sigma(E)) + (1 - f).
\end{math}  The parameter $\sigma$(E) is the photoelectric cross section.  Upon substituting in this model, we find that the covering fraction is not well constrained, with a range of possible values from 0.0--0.96.  Additionally, there is no change in $\chi^2$ when the {\tt pcfabs} model is used in place of the {\tt ztbabs} model.

\begin{figure}
\centering
\includegraphics[width=18cm]{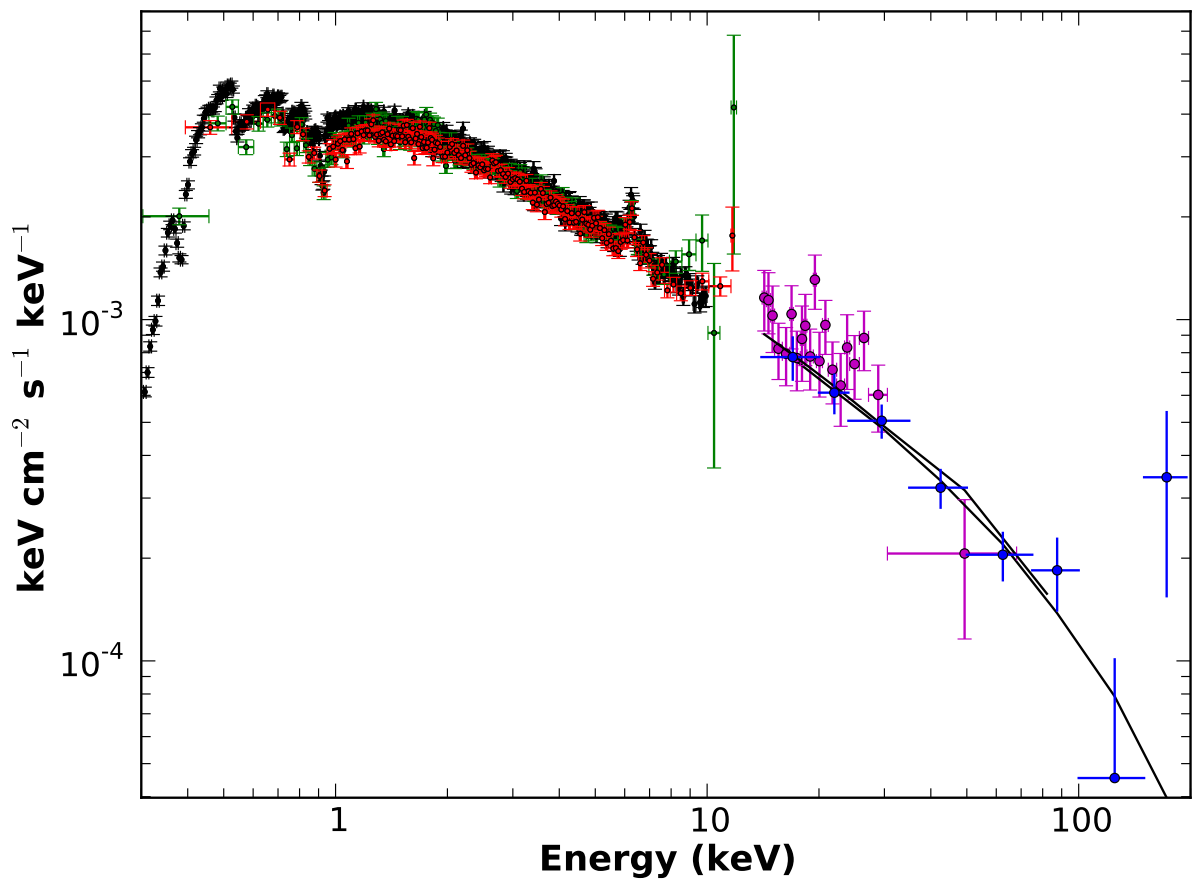}
\caption{Plotted is the unfolded spectrum for the broadband XMM-Newton pn, Suzaku XIS and PIN, and Swift BAT spectral fits.  Our best fit model includes edges and Gaussian lines to fit the prominent warm absorption features (\ion{O}{7}, \ion{O}{8}, \ion{N}{7}, and \ion{Ne}{9}), a black body to fit the soft excess, a neutral column of N$_{\rm H} \approx 10^{21}$cm$^{-2}$, a broad ($\sigma = 0.12$\,keV) Fe K$\alpha$ line, and a cutoff power law.  Best-fit parameters for this model are shown in Table~\ref{tbl-broad}.
\label{fig-broadband}}
\end{figure}

Adding an energy cutoff model to the power law component improves the fit significantly ($\Delta\chi^2 = -25$).  In this model ({\tt cutoffpl}), the power law component is multiplied by a factor of
\begin{math}
M(E) = \exp(-E/E_c),
\end{math}
where $E_c$ is the cutoff energy.  The best fit cutoff energy is 103$^{+72}_{-31}$\,keV, which is consistent with a Comptonization spectrum.  Best fit parameters for this full model (in XSPEC, the model is {\tt ztbabs}*{\tt zedge}*{\tt zedge}*((3 {\tt zgauss} components) + {\tt bbody} + {\tt cutoffpl})*{\tt const}) are shown in Table~\ref{tbl-broad} (we do not include the parameters for the soft energy edges and lines, instead see Table~\ref{tbl-softepic}).  Additionally, we plot the best-fit spectrum in Figure~\ref{fig-broadband}.

To determine the importance of reflection in the spectrum of NGC 6860, we included reflection with the XSPEC {\tt pexrav} model for Compton reflection off neutral material \citep{1995MNRAS.273..837M}.  We fixed the normalization and power law component to be the same between the reflection and cutoff power law models.  We also fixed the folding energy of the reflection component to be the same as the power law cutoff energy.  In the {\tt pexrav} model, we constrain the inclination of the disk to be $45\degr$ and the Fe abundance to be the solar value.  The best-fit reflection model is statistically worse than the cutoff power law model ($\Delta\chi^2 = 40$), but still an acceptable fit to the data with a reduced $\chi^2$ value of $1.11$.  We find that the best-fit reflection parameter ($R = \Omega / 2\pi$, where $\Omega$ is the angle subtended by the reflector) is $R = 1.15^{+0.46}_{-0.48}$.  With this reflection model, the best-fit neutral column density is slightly lower ($N_{\rm H} = 6.18^{+0.62}_{-0.89} \times 10^{20}$\,cm$^{-2}$).  The power law index ($\Gamma = 1.66^{+0.03}_{-0.04}$) and cutoff energy ($E_{cutoff} = 56^{+27}_{-15}$\,keV) are slightly different than in the direct power law model.  The relative normalizations from this model (relative to the Suzaku XIS spectra) are the pn at 1.09, the BAT at 0.94, and the PIN again fixed at 1.16.

\subsection{Better Constraining the Properties of the Warm Absorber}~\label{better}
Given the small band pass for the RGS spectra, the AGN continuum emission can not be constrained without the use of higher energy data.  In our earlier fits of the RGS spectra, we found that the fitted power law continuum was much steeper than the pn in a similar soft bandpass.  The value from the soft band pn continuum fits was also not well-constrained -- with a spectral index much flatter than that of the broad-band continuum fits.  Thus, to develop the best constraints on the warm absorption, we used the best-fit power law value from the broad-band fits along with the high resolution RGS spectra.  We found it necessary to fix these power law parameters in the model fits, due to degeneracies between the power law, soft excess, and absorption models in this narrow bandpass.

In the spectral analysis of a similar warm absorber source, NGC 3227, \citet{2009ApJ...691..922M}
compared three different models to the soft excess in a broad band EPIC spectral fit: a steep power law, a blackbody, and a Comptonized component (using the {\tt CompST} model).  The best statistical fit was with the Comptonization model (though it was only slightly better than a steep power law model).  Therefore, we chose to use this model in our new RGS spectral fit.  The values of the parameters from our best-fit model for the Comptonization parameter are a temperature of $kT = 1.09$\,keV, an optical depth of $\tau = 12$, and a normalization of $1.7 \times 10^{-3}$ ph\,keV$^{-1}$\,cm$^{-2}$\,s$^{-1}$.  We could not constrain the errors on this model as the parameters of this Comptonization model were not well-constrained.

The final model for this RGS spectral fit includes neutral absorption and two warm absorber components.  To constrain the velocity of the outflow, we allowed the redshift of the warm absorber components to vary.  However, we kept the turbulent velocity fixed at the default value of 100\,km\,s$^{-1}$.  We include the best-fit absorption properties for this model in Table~\ref{tbl-rgswarm}.  Both the measured column densities of the cold neutral and warm absorbers and the ionization parameters are well matched between the previous model fits in Table~\ref{tbl-warmabs2} and the new values in Table~\ref{tbl-rgswarm}.  However, we have placed tighter constraints on the model in our new fits by using a fixed power law component as determined from the broad-band fits.  Also, we have placed estimates on the velocity of the outflow.  The associated systemic velocities of both components (calculated from the redshifts recorded in Table~\ref{tbl-rgswarm}) are $v_1 = -184(^{+371}_{-110})$ km\,s$^{-1}$ and $v_2 = -196(^{+135}_{-100})$.  This places constraints on the outflowing gas to be from about 0--300\,km\,s$^{-1}$.

\section{The Unusual Properties of NGC 6860}~\label{unusual}
In \citet{2008ApJ...674..686W}, we pointed out the unusual spectral properties of NGC 6860 based on a $\approx 10$\,ks XMM-Newton follow-up observation.  The flat ($\Gamma \approx 1.0$) spectrum at hard energies led us to classify the source as a Compton thick candidate.  Compton thick sources are those where the optical depth towards Compton scattering is $\tau >> 1$, corresponding to column densities $N_{\rm H} \ga 10^{24}$\,cm$^{-1}$.  Compton thick spectra are reflection dominated, since little to no direct emission escapes below 10\,keV.  Thus, both a flat spectrum and high EW Fe K$\alpha$ emission line -- signatures of a reflection spectrum -- are diagnostics used in classification of a source as Compton thick.  Another possible diagnostic is a lack of variability, as discussed in \citet{2009ApJ...701.1644W}.

We classified the short XMM-Newton observation as a candidate Compton-thick source due to the unusual properties of the spectrum.  While the source showed a flat spectrum at high energies, the measured column density indicated a column density of only $\approx 10^{22}$\,cm$^{-2}$.  Additionally, the flat spectral index measured at low energies ($E = 0.3$--2\,keV), suggested that the source spectrum also includes a contribution from warm ionized gas.  Without a higher signal-to-noise observation we could not further determine the X-ray spectral properties of NGC 6860.

From our analysis of the new higher signal-to-noise spectra presented in this paper, we find that the spectra of NGC 6860 do not meet the criteria typically used to classify a source as Compton thick.  The continuum at high energies is no longer flat -- marking a clear change in spectral state from the earlier XMM-Newton spectrum.  Additionally, the Fe K$\alpha$ line has an equivalent width too low ($EW \approx 100$\,eV) to suggest a reflection dominated spectrum.  Further, the measured column density is well below the Compton thick threshold.  Therefore, the source is most likely in a Compton thin state throughout both the Suzaku and long XMM-Newton observations.  Still, it is unusual to find a Compton thin source whose flux and spectral shape remain constant over a period of a year, as we have found from our comparison of the Suzaku and XMM-Newton spectra.  Further yet, the Swift BAT 54-month light curve shows no variability at the highest energies over the entire 4.5\,year period (private communication, Gerry Skinner).


This lack of variability is surprising considering that the previous analysis of the short XMM-Newton and Swift XRT spectra showed variability in flux and spectral shape between the two observations spaced only 4 months apart.  In order to compare the spectra from the high (long exposure XMM-Newton) and low (short exposure XMM-Newton and Swift XRT) signal-to-noise observations, we fit the 0.3--10\,keV band of the new EPIC spectra with the partial covering absorbed power law model used in  \citet{2008ApJ...674..686W}.  This fit is not a good representation of the high quality EPIC spectra (with a reduced $\chi^2$ value of 3.4), but it allows us to compare the fitted parameters and flux between the old XMM, XRT, and new XMM observations.  The best-fit parameters to the new data are: N$_{\rm H} = 2.7 \times 10^{21}$\,cm$^{-2}$, with a covering fraction of 43\%, and $\Gamma = 1.58$.  As stated earlier, the spectral shape is much different than the old XMM observation from 3 years earlier (N$_{\rm H} \approx 4.5 \times 10^{22}$\,cm$^{-2}$, with a covering fraction of $< 70$\%, and $\Gamma \approx 1.0$), but similar to the values from the XRT observation.  Unfortunately, without a higher quality spectrum during the spectral state exhibited in the earlier XMM-Newton observation, we can not confirm that the source changed from a Compton thick to Compton thin phase.

Another noticeable change between the short XMM-Newton and XRT observations and the later Suzaku and long XMM-Newton observation is the change in flux.  The Suzaku and long XMM-Newton observations are 3 times brighter than the short XMM-Newton observation and 60\% brighter than the Swift XRT observation.  It is unclear why the more recent observations do not change -- whether we happened to catch the source in a similar state or the source somehow changed such that the X-ray spectrum is now constant.  However, given the history of past variability as well as the low Fe K$\alpha$ EW, it is unlikely that this object is truly Compton thick.  Although, it is possible that the source changed from a Compton thick state -- as is the case for the `changing-look' AGN \citep{2003MNRAS.342..422M}.  

When we examined the properties of the warm absorbers in NGC 6860, we find more evidence that its spectrum is unusual.  We find that there are likely absorption and emission components in the spectrum, but the only strong, clearly identifiable features at soft energies are the Fe M-shell UTA and the \ion{Ne}{9} absorption line.  We are unaware of any other warm absorber spectra that show \ion{Ne}{9} as the most prominent line in the soft spectrum.  While other absorption and emission features are present, it is impossible to clearly separate the features which are likely blended at the resolution available in the long XMM-Newton spectrum.

One of the clearest indications of this complexity is the discrepancy we get between the estimated velocities of the edges and lines identified in the PN and RGS spectra.  We find, for instance, that 
the fitted energy of the \ion{N}{7} absorption line shows no velocity shift, while the \ion{N}{6} and \ion{O}{7} lines are shifted by an apparent blue-shifted velocity of about 7000\,km\,s$^{-1}$.  To add to the complexity, we find that the \ion{O}{7} and \ion{O}{8} edges may either not be shifted or blue-shifted depending on the continuum and fitted energies of nearby lines.  Even in the RGS, we appear to have insufficient resolution to clearly identify all of the features.  Further, the warm absorption models, while they give an adequate fit, clearly do not fully account for all of the features we find (for instance, the \ion{Ne}{9} line has a higher EW than that fitted by the warm absorption models).  In order to get a better understanding, particularly of the kinematics of the gas, it will be imperative to obtain a UV spectrum of NGC 6860.

\begin{figure}
\centering
\includegraphics[width=18cm]{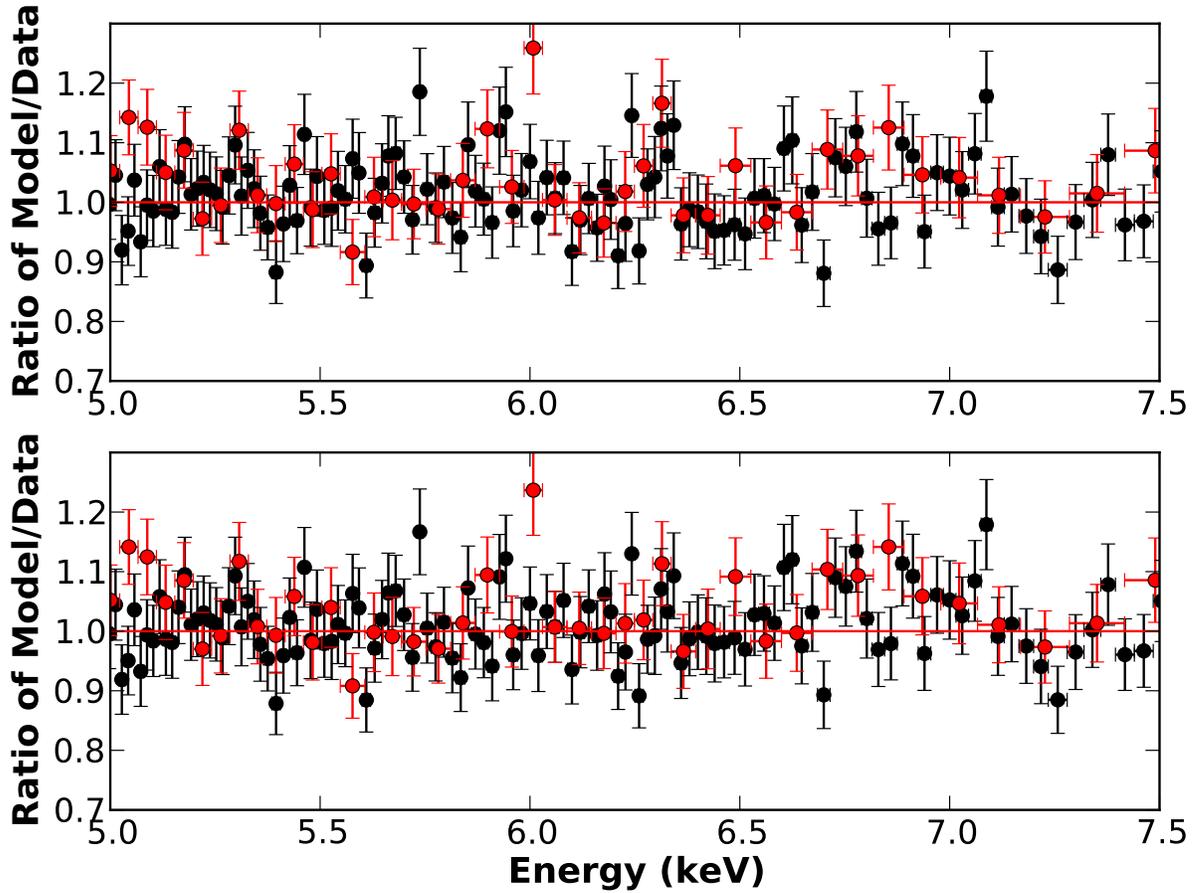}
\caption{Plotted is the ratio of the data/model of the Fe K$\alpha$ region using the Gaussian model (top panel) and a diskline model (bottom panel).  Both the XMM-Newton pn (black) and Suzaku XIS0+XIS3 (red) spectra are shown.  There is very little difference in the ratio plots from both models.  Statistically, the reduced $\chi^2$ is slightly lower with the Gaussian model.
\label{fig-comparefe}}
\end{figure}

Finally, another unusual property of the NGC 6860 spectra involves the fluorescent Fe line in the hard spectrum.  The average measured energy for the 6.4\,keV Fe K$\alpha$ line from Chandra grating observations of a sample of AGNs is $6.399 \pm 0.003$\,keV \citep{2004ApJ...604...63Y}.  We, however, measure a central energy of $6.331^{+0.030}_{-0.022}$, or red-shifted by about 2800\,km\,s$^{-1}$.  To confirm the measured energy of this line, we replaced the broad Gaussian model initially used to fit Fe K$\alpha$ in the broad-band spectral fit (Table~\ref{tbl-broad}) with a more physical model of line emission from a relativistic accretion disk (using the {\tt diskline} model of \citet{1989MNRAS.238..729F}).  In this model, we used the default emissivity index ($\beta = -2$) and outer radius (1000 GM/c$^2$).  We allowed the line energy, disk inner radius, and inclination to vary.  The best-fit diskline model, with $\chi^2$/dof $= 4282.4$/3872,  is statistically a worse fit ($\Delta\chi^2 \approx 28$) than the simple Gaussian.  However, there is little noticeable difference in the ratio of the data/model between the Gaussian and diskline models, as shown in Figure~\ref{fig-comparefe}.  The best-fit from the diskline model confirms the energy measured from the Gaussian model, with $E = 6.327^{+0.016}_{-0.017}$\,keV.  Additional fitted parameters for the diskline model are an inner accretion disk radius of $\le 23$ GM/c$^2$, inclination of $\le 15\degr$, and a normalization of $2.4^{+0.4}_{-0.6} \times 10^{-5}$\,photons\,cm$^{-2}$\,s$^{-1}$ (with a line equivalent width of 98\,eV).  With the diskline model, however, the 5.95\,keV line is no longer significant in the spectrum.

\section{Summary}\label{summary}
Following upon the analysis of a short $\approx 10$\,ks XMM-Newton observation of the Swift BAT-detected Sy 1.5 NGC 6860, we obtained and analyzed higher signal-to-noise observations of this source with Suzaku ($\approx 40$\,ks) and XMM-Newton ($\approx 100$\,ks).  Though the new observations show NGC 6860 in a different spectral state -- with a higher flux and steeper power law index -- these higher quality spectra confirm the complex nature of the X-ray spectrum of NGC 6860.  Even with the high signal-to-noise in these observations, we find that the X-ray spectra are still too complex to completely characterize.

Still, a number of conclusions can be drawn from our analysis.  Among these, we find that while short term variability is observed in both the Suzaku and XMM-Newton CCD spectra, both the flux and spectral shape remain similar between these observations though they were taken a year apart.  The overall spectral shape of NGC 6860 shows clear signs of a warm ionized absorber at soft energies along with a soft excess.  While a combination of two warm absorber components and a neutral absorber along with a base blackbody + power law model provides an acceptable spectral fit in the lower signal-to-noise of the Suzaku XIS spectra, the higher signal-to-noise/resolution XMM-Newton pn and RGS spectra are more difficult to describe. Particularly, while the Fe UTA features are well fit with the warm absorber models, we find additional absorption features (like the \ion{N}{7} absorption line) which are not well-fit by this warm absorber model below $\approx 0.7$\,keV.  

The properties of the warm absorbers in NGC 6860 are consistent with those of other Sy 1s, as determined through analyses of grating spectra \citep{2005AA...431..111B,2007MNRAS.379.1359M}. For instance, \citet{2007MNRAS.379.1359M} found in a sample of 15 AGN with Chandra HETGS spectra, that type 1 AGN typically have multiple warm absorber components with ionizing columns from $10^{20}$--$10^{23}$\,cm$^{-2}$, ionization parameters of $\xi \approx  10^0$--$10^4$, and velocities from $\approx 0$--$2000$\,km\,s$^{-1}$.  Our analysis of NGC 6860 indicates two components with ionized columns of $\approx 10^{20}$--$10^{21}$\,cm$^{-2}$, with ionization parameters of $\xi \approx 45$ and $180$\,ergs\,s$^{-1}$, and outflowing velocities between 0 and 300\,km\,s$^{-1}$.  However, the constraints on the outflow velocities are uncertain since measurements of individual lines uncover different central energies depending on the model used.  Thus, UV data will be particularly important to obtain for NGC 6860 to determine the kinematic properties of the outflowing material.   

We find complexity not only in the soft spectrum but also in the hard X-ray spectrum of NGC 6860.  
The hard spectra show a broad ($\approx 11000$\,km\,s$^{-1}$) Fe\,K$\alpha$ line with $EW \approx 100$ in both the XMM-Newton EPIC and Suzaku XIS observations.  The FWHM of the iron line is much higher than the measured FWHM for H$\beta$ ($5920 \pm 600$\,km\,s$^{-1}$; 
\citet{2006AA...459...55B}), indicating that the Fe K$\alpha$ line originates in a region interior to the broad line region and closer to the central black hole.  The energy of this line is low ($6.35 \pm 0.02$ keV), suggesting a red-shift in the 6.41\,keV \ion{Fe}{1} K$\alpha$ line corresponding to $v \approx 2800$\,km\,s$^{-1}$.  This energy is unusual, considering that typical measurements from Chandra high energy grating spectra have mean line-center energies of $6.399 \pm 0.003$\,keV \citep{2004ApJ...604...63Y}.
  An additional possibly redshifted narrow Fe K emission line is significant in the XMM-Newton spectra with an energy of $5.95^{+0.03}_{-0.06}$\,keV ($v \approx 0.08c$) and $EW = 17$\,eV.  However, with the more physical diskline model, the 5.95\,keV line is no longer significant.

The shape of the broad band continuum -- measured from a combination of XMM-Newton, Suzaku, and Swift BAT observations -- is well fit by an absorbed power law model.  The column density of the neutral absorber is $\approx 10^{21}$\,cm$^{-2}$.  The power law index ($\Gamma = 1.64$) is lower than the typical value of 1.75 measured for the Swift BAT AGN sample \citep{2009ApJ...690.1322W}, but much higher than the value of 1.0 measured in the earlier short XMM-Newton observation of this source.  The high energy cutoff for the power law is about 100\,keV, which is consistent with a Comptonization spectrum.

\acknowledgments
The authors gratefully acknowledge the work of the Swift BAT team, in particular Jack Tueller, Wayne Baumgartner, and Gerry Skinner, as well as Tim Kallman for his help with using the XSPEC {\tt warmabs} model.  This research has made use of data obtained from the Suzaku satellite, a collaborative mission between the space agencies of Japan (JAXA) and the USA (NASA). L.M.W. acknowledges support through NASA grant HST-HF-51263.01-A, through a Hubble Fellowship from the Space Telescope Science Institute, which is operated by the Association of Universities for Research in Astronomy, Incorporated, under NASA contract NAS5-26555.

{\it Facilities:} \facility{Swift ()}, \facility{Suzaku ()}, \facility{XMM-Newton ()}

\bibliography{ms.bib}

\clearpage
\begin{deluxetable}{lcccc}
\tablecaption{Features Detected in the Soft XMM-Newton EPIC spectra\label{tbl-softepic}}
\tablewidth{0pt}
\tablehead{
\colhead{Component} & \colhead{Energy (keV)} &
\colhead{$\tau$}  & \colhead{$\sigma$ (keV)} & \colhead{$I$ (photons\,cm$^{-2}$\,s$^{-1}$)} 
}

\startdata
\ion{O}{7} edge & $0.719^{+0.004}_{-0.004}$ & $0.178^{+0.013}_{-0.007}$ & -- & -- \\
\ion{O}{8} edge & $0.845^{+0.004}_{-0.003}$ & $0.254^{+0.005}_{-0.019}$ & -- & -- \\
\ion{Mg}{11} edge & $1.717^{+0.039}_{-0.023}$ & $0.027^{+0.006}_{-0.008}$ & -- & --  \\
\ion{N}{6} line & $0.405^{+0.006}_{-0.405}$ & -- & 0.01 & $-4.47^{+0.37}_{-0.57} \times 10^{-4}$\\
\ion{N}{7} line (?) & $0.451^{+0.006}_{-0.004}$ & -- &  $0.01^{+0.176}$ & $2.45^{+3.84}_{-0.26} \times 10^{-4}$\\
\ion{Ne}{9} line & $0.937^{+0.007}_{-0.004}$ & -- & 0.01 & $-4.53^{+0.44}_{-0.69} \times 10^{-5}$\\
\ion{Si}{13} line & $1.829^{+0.012}_{-0.011}$ & -- & 0.01 & $-1.37^{+0.25}_{-0.23} \times 10^{-5}$\\
\enddata
\par Listed are additional features that significantly ($|\Delta\chi^2| > 15$) improved the fit to the EPIC pn and MOS spectra.  These features were added in addition to a power law absorbed by the Galactic column in the 0.3--3\,keV band.  For the MOS spectra, we excluded data $< 0.5$\,keV, due to inconsistencies with the pn spectra.  Negative intensities indicate absorption lines, while positive intensities indicate emission lines.  The recorded energies are in the rest frame of the AGN.
\end{deluxetable}

\begin{deluxetable}{cclc}
\tablecaption{Lines Detected in the RGS spectra\label{tbl-rgs}}
\tablewidth{0pt}
\tablehead{
\colhead{Energy (eV)} & 
\colhead{$I (\times 10^{-4}$\,photons\,cm$^{-2}$\,s$^{-1}$)}  & \colhead{ID} &
\colhead{$|\Delta\chi^2|$}
}

\startdata
411$^{+18}_{-31}$ & --1.38$^{+0.48}_{-1.52}$ & \ion{N}{6} & 17.7\\
502$^{+8}_{-8}$ & 1.17$^{+0.25}_{-0.24}$ & \ion{N}{7} & 48.7\\
545$^{+4}_{-4}$ & --0.77$^{+0.32}_{-0.31}$ & \ion{O}{7} & 82.1\\
917$^{+7}_{-7}$ & --1.56$^{+0.22}_{-0.26}$ & \ion{Ne}{9} & 163.5\\
\enddata
\par Listed are the fitted energy, intensity, probable identification, and change in $\chi^2$ upon adding a Gaussian to model the indicated line at the redshift of the source ($z = 0.014884$).
\end{deluxetable}

\begin{deluxetable}{llll}
\tablecaption{Spectral Fits in the 0.3--3\,keV band with Warm Absorption Models\label{tbl-warmabs2}}
\tablewidth{0pt}
\tablehead{
\colhead{Component} & \colhead{Suzaku XIS} & \colhead{XMM PN} &
\colhead{XMM RGS}
}

\startdata
N$_H$\tablenotemark{a} (cold) & 10.91$^{+2.96}_{-0.94}$
& 6.41$^{+0.04}_{-0.08}$ & $9.79^{+0.66}_{-1.76}$ \\
\hline
N$_H$\tablenotemark{a} (warm 1) & 7.40$^{+5.16}_{-3.41}$ 
& 12.52$^{+0.83}_{-0.90}$ & 14.16$^{+7.33}_{-4.85}$ \\
$\log \xi$\tablenotemark{b} (warm 1)& 1.95$^{+0.02}_{-0.09}$ & 2.04$^{+0.04}_{-0.01}$ & 2.42$^{+0.05}_{-0.13}$ \\
v$_{turb}$\tablenotemark{b} (warm 1)& 52$^{+30}_{-16}$ & 25$^{+6}_{-5}$ 
& 83$^{+32}_{-26}$ \\
\hline
N$_H$\tablenotemark{a} (warm 2)& 7.72$^{+1.49}_{-2.55}$
& -- & 0.59$^{+0.15}_{-0.14}$\\
$\log \xi$\tablenotemark{b} (warm 2) &  2.65$^{+0.07}_{-0.07}$& -- & 1.64$^{+0.07}_{-0.07}$\\
v$_{turb}$\tablenotemark{b} (warm 2) & 185$^{+116}_{-77}$ & -- &  113\tablenotemark{\dagger}\\
\hline
$kT$ (keV) & 0.102$^{+0.022}_{-0.022}$ & 0.131$^{+0.002}_{-0.003}$ & 0.87\tablenotemark{\dagger} \\
$I_{kT}$\tablenotemark{c} & 0.47$^{+0.09}_{-0.14}$ & 0.27$^{+0.01}_{-0.01}$& 1.26$^{+36.45}_{-0.39}$\\
$\Gamma$ & 1.74$^{+0.04}_{-0.03}$ & 1.55$^{+0.01}_{-0.01}$ & 2.40$^{+0.18}_{-0.22}$\\
$I_{\Gamma}$\tablenotemark{c} & 64.46$^{+0.84}_{-3.61}$ & 64.88$^{+0.48}_{-1.18}$ & 60.32$^{+6.07}_{-4.28}$\\
$\chi^2/{\rm dof}$ & 938.9/875 & 727.0/534 & 2444.4/1951 \\

\enddata
\tablenotetext{a}{Column density of the neutral ({\tt ztbabs}) or warm absorber ({\tt warmabs}) in units of $10^{20}$\,cm$^{-2}$.\\}
\tablenotetext{b}{Additional components of the warm absorber model include the ionization parameter ($\log \xi$, where $\xi$ has units of ergs\,s$^{-1}$) and the turbulent velocity (in units of km\,s$^{-1}$).\\}
\tablenotetext{c}{Intensity of the model component in units of $10^{-4}$ photons\,cm$^{-2}$\,s$^{-1}$ at 1\,keV for either the blackbody component (kT) or the power law ($\Gamma$) component.\\}
\tablenotetext{\dagger}{The indicated parameter is an upper limit.}
\end{deluxetable}

\begin{deluxetable}{ll}
\tablecaption{Broad Band Spectral Fits in the 0.3--195\,keV Range\label{tbl-broad}}
\tablewidth{0pt}
\tablehead{
\colhead{Component} & \colhead{Fitted Value}
}
\startdata
N$_{\rm H}$\tablenotemark{a} (cold) & 9.96$^{+1.64}_{-1.17}$\\
kT (keV) & 0.067$^{+0.009}_{-0.009}$\\
I$_{kT}$\tablenotemark{b} & 0.74$^{+0.22}_{-0.17}$\\
$\Gamma$ & 1.64$^{+0.02}_{-0.02}$\\
$E_{cutoff}$ (keV) & 103$^{+72}_{-31}$\\
I$_{\Gamma}$\tablenotemark{b} & 51.10$^{+0.67}_{-0.69}$\\
 E$_{Fe K\alpha}$ (keV) & 6.331$^{+0.030}_{-0.022}$\\
$\sigma_{Fe K\alpha}$ (keV) & 0.120$^{+0.033}_{-0.026}$\\
EW$_{Fe K\alpha}$ (eV) & 86.8$^{+14.2}_{-14.3}$\\
I$_{Fe K\alpha}$\tablenotemark{b} & 0.22$^{+0.04}_{-0.04}$\\
$\chi^2/$dof & 4254.3/3873\\
\\
F$_{0.3-10 keV}$ (pn) & $3.4 \times 10^{-11}$\,ergs\,s\,cm$^{-2}$ \\
F$_{15-50 keV}$ (PIN) & $3.3 \times 10^{-11}$\,ergs\,s\,cm$^{-2}$ \\
F$_{14-195 keV}$ (BAT) & $4.2 \times 10^{-11}$\,ergs\,s\,cm$^{-2}$ \\
\enddata
\tablenotetext{a}{Column density of the neutral absorber from the {\tt ztbabs} model in units of $10^{20}$\,cm$^{-2}$.\\}
\tablenotetext{b}{Intensity of the model component in units of $10^{-4}$ photons\,cm$^{-2}$\,s$^{-1}$ at 1\,keV for either the blackbody component (kT), the power law ($\Gamma$) component, or
a Gaussian component.\\}
\end{deluxetable}

\begin{deluxetable}{ll}
\tablecaption{RGS Best-fit Absorption Values\label{tbl-rgswarm}}
\tablewidth{0pt}
\tablehead{
\colhead{Parameter} & \colhead{Fitted Value}
}
\startdata
N$_{\rm H}$\tablenotemark{a} (cold) & 9.24$^{+0.90}_{-0.71}$\\
\hline
N$_{\rm H}$\tablenotemark{a} (warm 1) & $6.90^{+1.73}_{-1.41}$\\
$\log \xi$\tablenotemark{b} (warm 1) & $2.26^{+0.24}_{-0.14}$\\
$z_{absorber}$ (warm 1) & $0.014269^{+0.00053}_{-0.00054}$\\
\hline
N$_{\rm H}$\tablenotemark{a} (warm 2) & $0.59^{+0.17}_{-0.12}$\\
$\log \xi$\tablenotemark{b}  (warm 2)& $1.65^{+0.05}_{-0.13}$\\
$z_{absorber}$ (warm 2) & $0.014231^{+0.000449}_{-0.000335}$\\ 
\hline
$\chi^2$/dof & 2458.8/1952\\
\enddata
\tablenotetext{a}{Column densities of the neutral and warm absorbers from the {\tt ztbabs} or {\tt warmabs} model are in units of $10^{20}$\,cm$^{-2}$.\\}
\tablenotetext{b}{The ionization parameter ($\log \xi$, where $\xi$ has units of ergs\,s$^{-1}$) is given from the warm absorption model.\\}
\end{deluxetable}

\clearpage

\end{document}